\documentclass[a4paper,10pt]{article}
\usepackage{graphicx}
\usepackage{amssymb}
\usepackage{amsmath}
\usepackage{amsfonts}
\usepackage{amsthm}

\usepackage{nicefrac}
\usepackage{dcolumn}
\usepackage{multirow}
\usepackage{cite}
\usepackage{mathrsfs}
\usepackage{bm}
\usepackage[usenames,dvipsnames]{xcolor}
\usepackage[colorlinks=true,citecolor=Blue,linkcolor=RubineRed,urlcolor=Blue]{hyperref}
\usepackage{bbm}
\usepackage{hyperref}
\usepackage{comment}
\usepackage{float}
\usepackage{caption}
\usepackage{subcaption}

\usepackage{xcolor}
\usepackage{soul}
\setstcolor{red}

\begin{document}

\huge

\begin{center}
Average-atom approach for transport properties of shocked argon in the presence of a magnetic field
\end{center}

\vspace{0.5cm}

\large

\begin{center}
Nadine Wetta$^{a,}$\footnote{nadine.wetta@cea.fr} and Jean-Christophe Pain$^{a,b}$
\end{center}

\normalsize

\begin{center}
\it $^a$CEA, DAM, DIF, F-91297 Arpajon, France\\
\it $^b$Universit\'e Paris-Saclay, CEA, Laboratoire Mati\`ere sous Conditions Extr\^emes,\\
\it F-91680 Bruy\`eres-le-Ch\^atel, France
\end{center}

\vspace{0.5cm}

\begin{abstract}
We present electron transport calculations of shocked argon based on an average-atom modeling of the plasma, and compare them with measurements, involving both incident and reflected shock waves. Since the corresponding experiments are subject to a 5 T magnetic field, the impact of the latter on the Rankine-Hugoniot equations is taken into account, starting from the magneto-resistive hydrodynamics, and the resistivity tensor is deduced from the Boltzmann equation. The resistivity tensor yields the electrical and Hall resistivities. Our average-atom code {\sc Paradisio} provides the quantities required for the calculation of electrical resistivity within the Ziman-Evans formalism, as well as for the Hall resistivity. We obtain a good agreement between calculated conductivities and experimental values, both for the incident and reflected shocks. Our values of the Hall constant are compared to experimental values derived from Hall voltage measurements, as well as to theoretical ones based on the quantum statistical linear-relaxation-time approach.
\end{abstract}

\vspace{0.5cm}

\section{Introduction}\label{sec1}

Argon is the most abundant noble gas on Earth, and is also present in the atmosphere of gaseous giant planets \cite{Atreya2003,Hersant2004,Mousis2010}. Understanding the physics of the latter requires accurate equations of state (EOS), as well as theoretical transport coefficients for this element. More generally, argon is also an ideal candidate for theoretical studies of warm-dense-matter (WDM), due to its high ionization energy (15.76 eV \cite{Kramida2024}, only surpassed by He, F and Ne ones). As a result, argon remains in partially ionized states over wide density and temperature ranges. Partial ionization is one of the main features of WDM, and the determination of the free-electron density $n_e$ is of particular importance under these conditions. Experiments are also essential as reference points for WDM theoretical models.

Experiments involving noble gas plasmas, including He, Ne, Ar, Kr and Xe, have been conducted over the last 30 years using explosively driven shock wave plasmas. Noble gas plasmas were created with temperatures $T$ of 6000-10$^5$ K and densities of 0.001 - 10 g/cm$^3$ \cite{Ivanov1976,Gatilov1985,Shilkin2003}. The dc conductivity was measured in each experiment.
In the most recent experiments by Shilkin \emph{et al.} \cite{Shilkin2003}, shock loading experiments have been realized in presence of magnetic fields for Xe and Ar, providing measurements of the Hall voltage. The latter are expected to be a more direct diagnostic tool for $n_e$ than electrical conductivity.

In order to properly describe transport properties of dense plasmas, a consistent quantum statistical description of electronic structure is necessary. Adams \emph{et al.}  used an approach based on linear-response-theory (LRT) as proposed by Zubarev \cite{Zubarev1996,Ropke2013}, to describe transport properties in terms of force-force correlation functions calculated within perturbation theory \cite{Adams2007,Adams2007b,Adams2007c,Adams2010}. This model enables for a complete description of partially ionized plasmas, accounting for electron interactions with other electrons, with different ionic species (carrying different charges $Z=1$, 2,\dots) and with neutral argon atoms. To the best of our knowledge, their work represents to date the most complete theoretical study  of the transport properties in argon under these very weak ionization conditions.

In this paper, we present an alternative study based on the use of average-atom method to derive a collision frequency between electrons and a mean ion, whose low charge is representative of the average charge carried by the actual various ion species. This approach avoids the difficulties of calculating the actual plasma composition and of modeling different collision frequencies between electrons and numerous species composing the plasma. Electron-ion collisions are easily obtained with the $T-$matrix formalism. As far as we know, there are few theoretical models for the electron-neutral collision frequency, under the conditions reached in the experiments carried by Shilkin \emph{et al.} on argon.

Average-atom models are commonly used for direct-current (dc) electrical resistivity calculations within the Ziman-Evans formalism (see for instance Refs.~\cite{Hansen2006,Sterne2007,Faussurier2019,Dharma-wardana2017,Wetta2020,Wetta2023}), which was extended to other electron-transfer coefficients for the fully-ionized, fully-degenerate hydrogen plasma \cite{Minoo1976}.

The main features of Ziman's formalism for conductivity are recalled in section \ref{sec2}, as well as the way the required quantities are derived from our average-atom code {\sc Paradisio}, also described within this section.

Section \ref{sec3} opens with a brief description of the shock loading experiments of Shilkin \emph{et al.} on argon. Starting from the magnetohydrodynamics (MHD) equations, the Rankine-Hugoniot (RH) relations in presence of a magnetic field are derived. The Rankine-Hugoniot equations relate the thermodynamical conditions behind the shock wave to the incident ones. In presence of an external magnetic field, the relation between an applied electric field and the induced electric current is tensorial. We recall the main steps leading to the resistivity tensor, starting from the Boltzmann equation. 

Section \ref{sec4} presents our numerical electrical conductivity and Hall resistivity calculations in the conditions of the experiments of Shilkin \emph{et al.}. Our results are compared to experimental values, and, concerning Hall effect, to other theoretical results.

In section \ref{sec5}, we focus on comparisons with the LRT Hall constant calculations of Adams \emph{et al.}. The relevance of average-atom methods for these calculations in partially ionized plasmas is discussed, before finally concluding this work by a brief summary (section \ref{sec6}).

\section{Calculation of electrical resistivity in the framework of the average-atom model}\label{sec2}

\subsection{The Ziman-Evans formulation}\label{subsec21}

The Ziman resistivity formula \cite{Ziman1961} reads
\begin{equation}\label{eta}
    \eta=-\dfrac{1}{3\pi {Z^*}^2 n_i} \dfrac{\hbar}{e^2} \int_0^\infty \dfrac{\partial f}{\partial \epsilon}(\epsilon,\mu^*) \mathcal{I}(\epsilon) d\epsilon,
\end{equation}
where $n_i$ is the ion density, $Z^*$ the mean ionic charge, $\mu^*$ the chemical potential. $f(\epsilon,\mu^*)$ is the Fermi-Dirac distribution function:
\begin{equation}\label{fd}
    f(\epsilon,\mu^*)=\dfrac{1}{1+e^{\beta(\epsilon-\mu^*)}},
\end{equation}
where $\beta=1/(k_BT)$, $k_B$ denoting the Boltzmann constant. To be consistent with the uniform electron gas (UEG) assumption underlying the Ziman theory, the chemical potential $\mu^*$ is given by
\begin{equation}\label{mu*}
    Z^* n_i=\dfrac{\sqrt{2}}{\pi^2 \beta^{3/2}}\mathcal{F}_{1/2}(\beta\mu^*),
\end{equation}
where 
\begin{equation}\label{Fermi_1sur2}
    \mathcal{F}_{1/2}(x)=\int_0^\infty dt \dfrac{t^{1/2}}{e^{t-x}+1}.
\end{equation}
defines the Fermi function of order 1/2. The function $\mathcal{I}(\epsilon)$ is related to the scattering cross-section $\Sigma(q)$ and to the ion-ion structure factor $S(q)$ by
\begin{equation}
    \mathcal{I}(\epsilon)=\int_0^{2k}q^3 S(q) \Sigma(q) dq,
\end{equation}
where $\vec{q}=\vec{k}^\prime-\vec{k}$ is the momentum transferred in the elastic scattering event, ({\it i.e.}, in which $|\vec{k}^\prime|=|\vec{k}|$). Introducing the scattering angle $\theta\equiv (\vec{k},\vec{k}^\prime)$, one has $q^2=2k^2 (1-\chi)$, where $\chi=\cos\theta$, and one gets then the following expression introducing the squared modulus of the scattering amplitude $|a(k,\chi)|^2$
\begin{equation}
    \mathcal{I}(\epsilon)=2k^4 \int_{-1}^1 S\left[k\sqrt{2(1-\chi)}\right]|a(k,\chi)|^2 (1-\chi) d\chi.
\end{equation}
$|a(k,\chi)|^2$ is provided by the $T-$matrix formalism of Evans \cite{Evans1973} which reads, in the relativistic formalism underlying our average-atom code {\sc Paradisio} \cite{Penicaud2009,Sterne2007}
\begin{align}\label{scattering}
    |a(k,\chi)|^2=\dfrac{1}{k^2}\left(\Big|\sum_\kappa |\kappa|e^{i\delta_\kappa(k)}\sin[\delta_\kappa(k)]P_{\ell}(\chi)\Big|^2 \right.\nonumber\\
    +\left.\Big|\sum_\kappa \dfrac{|\kappa|}{i\kappa}e^{i\delta_\kappa(k)}\sin[\delta_\kappa(k)]P^1_\ell(\chi)\Big|^2\right),
\end{align}
where $\kappa=-(\ell+1)$ for $j=\ell+1/2$, $\kappa=\ell$ for $j=\ell-1/2$, $\ell$ being the usual orbital quantum number. $P_\ell$ and $P^1_\ell$ are the Legendre and associated Legendre polynomials. Finally, the quantities $\delta_\kappa(k)$ denote the scattering phase-shifts.\\

In the present work, the mean ionic charge $Z^*$ and the scattering phase-shifts $\delta_\kappa(k)$ are provided by the average-atom code {\sc Paradisio}. Outputs from this code are also used to build the ion-ion structure factor $S(q)$.\\

\subsection{The average-atom model \sc{Paradisio}}\label{subsec22}

Atomic units where $e=\hbar=m_e=1$, and where the velocity of light $c=137.036$ is the inverse of the fine structure constant $\alpha=e^2/(8\pi\epsilon_0 a_B)$, $a_B$ being the Bohr radius and $\epsilon_0$ the permittivity of vacuum, are used throughout this section.\\

The {\sc Paradisio} \cite{Penicaud2009} code is based on Liberman's relativistic quantum-average-atom model {\sc Inferno} \cite{Liberman1979} which considers the atom as a point nucleus surrounded by its $Z$ electrons, placed at the center of a spherical cavity of radius $R_\mathrm{ws}$ dug into a jellium. The Wigner-Seitz radius $R_\mathrm{ws}$ reads
\begin{equation}\label{Rws}
    R_\mathrm{ws}=\left(\dfrac{3}{4\pi}\dfrac{A/\mathcal{N}_\mathrm{Avo}}{\rho} \right)^{1/3},
\end{equation}
$\rho$, $A$ and $\mathcal{N}_\mathrm{Avo}$ denoting respectively the mass density, molar mass and Avogadro number.\\
The jellium consists in a uniform electron gas and a uniform distribution of positive charges that ensures its electrical neutrality. {\sc Inferno} model also imposes electrical neutrality inside the cavity. The electronic structure is computed in a self-consistent way. The only required parameters are atomic number $Z$, molar mass $A$, mass density $\rho$ and temperature $T$.\\
In this spherical symmetry, the one-electron wave-functions, solutions of Dirac equation, are of the form
\begin{equation}\label{psis}
    \psi_s(\vec{r})\equiv\psi_{j\ell m}(\vec{r})=\left(\begin{array}{l}
    \displaystyle\dfrac{1}{r}F(r)\Omega_{j\ell m}(\theta,\phi)\\
    -\displaystyle\dfrac{i}{r}G(r)\Omega_{j\ell'm}(\theta,\phi)
    \end{array}
    \right),
\end{equation}
where $\Omega_{j\ell m}$ and $\Omega_{j'\ell m}$ are two spinors. $j$, $\ell$ and $m$ are quantum numbers associated respectively to the total angular momentum $J$, to the orbital angular momentum $L$ and its projection $L_z$ on the $z$ axis. The quantum number $\ell'$ is given by: 
\begin{equation}
    \ell'=\left\{\begin{array}{ll}
    \ell+1\;\;\;\;\mathrm{if}\;\;\;\; j=\ell+1/2\\
    \ell-1\;\;\;\;\mathrm{if}\;\;\;\; j=\ell-1/2.
    \end{array}
    \right.
\end{equation}
The Dirac equation then reduces to the following equations satisfied by the radial functions $F(r)$ and $G(r)$ 
\begin{equation}\label{radial}
    \left\{
    \begin{array}{l}
    \dfrac{dF}{dr}=-\dfrac{\kappa}{r}F(r)-\dfrac{V_{\mathrm{eff}}(r)-c^2-\epsilon}{c}G(r)\\
    \\
    \dfrac{dG}{dr}=\dfrac{V_{\mathrm{eff}}(r)+c^2-\epsilon}{c}F(r)+\dfrac{\kappa}{r}G(r)
    \end{array}
    \right.
\end{equation}
where
\begin{equation}
    \left\{
    \begin{array}{l}
    \kappa=-(\ell+1)\;\;\;\;\mathrm{for}\;\;\;\; j=\ell+1/2,\\
    \kappa=\ell\;\;\;\;\;\;\;\;\;\;\;\;\;\;\;\;\mathrm{for}\;\;\;\; j=\ell-1/2.
    \end{array}
    \right.
\end{equation}
Outside the cavity, the effective potential $V_\mathrm{eff}(r)$ is constant and given by:
\begin{equation}
    V_{\infty}=\mu_{\mathrm{xc}}[\bar{n},T],
\end{equation}
$\bar{n}$ denoting the density of the jellium
\begin{equation}
    \overline{n}=\dfrac{\sqrt{2}}{\pi^2 \beta^{3/2}}\mathcal{F}_{1/2}(\beta\mu),
\end{equation}
with the Fermi function $\mathcal{F}_{1/2}(\beta\mu)$ expression given in Eq.~(\ref{Fermi_1sur2}). $\mu_\mathrm{xc}[\overline{n},T]$ is the electron exchange-correlation potential functional evaluated at the UEG density $\overline{n}$ and at temperature $T$. {\sc Paradisio} uses the finite temperature functionals of the KSDT form \cite{Karasiev2014} with revised parameters from Groth \emph{et al.} \cite{Groth2017}.\\

The model imposes $F(r)=G(r)=0$ at $r=0$ and $r\rightarrow\infty$. Outside the cavity, the radial functions $F^\mathrm{oc}(r)$ and $G^\mathrm{oc}(r)$ (the superscript ``oc'' stands for ``outside cavity'') satisfying those boundary conditions are, for bound states, (\emph{i.e.}, for energies $\epsilon<V_\infty$), modified Bessel functions of the third kind \cite{Abramowitz1965}, exponentially decreasing: 
\begin{equation}
    \epsilon<V_\infty: \hspace{1cm} \left\{
    \begin{array}{l}
    F^\mathrm{oc}_\mathrm{b}(r)=a_0\displaystyle\dfrac{kc}{V_{\infty}-\epsilon}rK_{\ell+1/2}(kr)\\
    \\
    G^\mathrm{oc}_\mathrm{b}(r)=a_0rK_{\ell'+1/2}(kr),
    \end{array}
    \right.
\end{equation}
and, for free states, (\emph{i.e.}, for energies $\epsilon\geq V_\infty$), combinations of Bessel functions of the first and second kinds, with decreasing amplitudes as $r\rightarrow\infty$:
\begin{gather}
    \epsilon\geq V_\infty: \hspace{1cm} 
    \left\{
    \begin{array}{l}
    F^\mathrm{oc}_\mathrm{f}(r)=b_0\displaystyle\dfrac{kc}{\epsilon-V_{\infty}}r\left[\cos(\delta_\kappa) j_{\ell}(kr)-\sin(\delta_\kappa) n_{\ell}(kr)\right]\\
    \\
    G^\mathrm{oc}_\mathrm{f}(r)=b_0r\left[\cos(\delta_\kappa) j_{\ell'}(kr)-\sin(\delta_\kappa) n_{\ell'}(kr)\right],\nonumber
    \end{array}
    \right.
\end{gather}
where $a_0$ and $b_0$ are two normalisation factors.
{\sc Paradisio} then only needs to solve Eqs.~(\ref{radial}) inside the cavity. The continuity condition at the cavity radius $r=R_\mathrm{ws}$ is only possible for discrete values of the energies $\epsilon<V_\infty$, yielding the bound states. The matching of inside and outside solutions is possible at any energy $\epsilon\geq V_\infty$ by adjusting the phase-shifts $\delta_\kappa(k)$, giving the continuum of free states.\\ 

The electronic density $n(r)$ is then obtained by
\begin{align}
    n(r)=&\sum_\mathrm{b} \sum_\kappa 2|\kappa|\,\left[F_\mathrm{b}(r,\kappa,\epsilon_\mathrm{b})^2+G_\mathrm{b}(r,\kappa,\epsilon_\mathrm{b})^2 \right]\nonumber\\
    &+\int_0^\infty d\epsilon\, \sum_\kappa 2|\kappa|\,\left[F_\mathrm{f}(r,\kappa,\epsilon)^2+G_\mathrm{f}(r,\kappa,\epsilon)^2 \right].
\end{align}
The number $Z_\mathrm{bound}$ of bound electrons and the number $Z_\mathrm{cont}$ of continuum ones respectively read
\begin{eqnarray}
    Z_{\mathrm{bound}}&=&\sum_ \mathrm{b} f(\epsilon_\mathrm{b},\mu) \sum_\kappa 2|\kappa|\\
    & &\times\left\{\int_0^{R_\mathrm{ws}}\left[F_\mathrm{b}(r,\kappa,\epsilon_\mathrm{b})^2+G_\mathrm{b}(r,\kappa,\epsilon_\mathrm{b})^2 \right]\,r^2 dr \right\},\nonumber
\end{eqnarray}
and
\begin{eqnarray}\label{zcont}
    Z_{\mathrm{cont}}&=&\int_0^\infty d\epsilon f(\epsilon,\mu) \sum_\kappa 2|\kappa|\\
    & &\times\left\{\int_0^{R_\mathrm{ws}}\left[F_\mathrm{f}(r,\kappa,\epsilon)^2+G_\mathrm{f}(r,\kappa,\epsilon)^2 \right]\,r^2 dr \right\}.\nonumber
\end{eqnarray}
The chemical potential $\mu$ is obtained from the charge neutrality condition $Z=Z_\mathrm{bound}+Z_\mathrm{cont}$ inside the cavity.

\subsection{Mean ionic charge and ion-ion structure factor from the code}

\begin{figure}[ht!]
\centering
\includegraphics[scale=0.5]{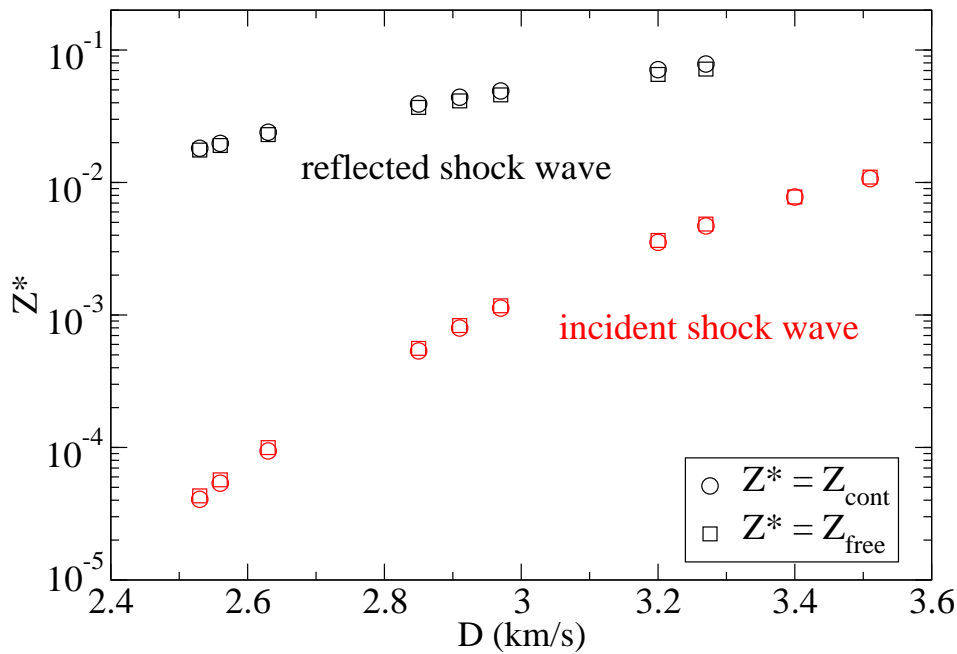}
\caption{\label{fig10} Argon mean ion charge $Z^*$ in the conditions of the shock experiments of Shilkin \emph{et al.}, from our average-atom code, using two widely used definitions. $Z^*=Z_\mathrm{cont}$, is the one based on the continuum density of states, giving the values represented by the circles.  The squares correspond to $Z^*=Z_\mathrm{free}$, \emph{i.e.}, the number deduced from the jellium's density of states. The data relative to the incident shock waves are presented in red, and those corresponding to the reflected shocks in black. $D$ denotes the shock speeds.}
\end{figure}

The scattering phase-shifts $\delta_\kappa(k)$ needed by the Ziman formalism are obtained from the continuity condition on the radial functions solutions of Eqs.~(\ref{radial}) at the cavity radius $r=R_\mathrm{ws}$ for $\epsilon\geq V_\infty$.\\

The formalism requires a value for the mean ionic charge $Z^*$. In the framework of our average-atom approach, the use of $Z_\mathrm{cont}$, given by Eq.~(\ref{zcont}) for this quantity seems obvious, as does that of the one obtained from the jellium's charge density, denoted $Z_\mathrm{free}$ and reading
\begin{equation}\label{zfree}
    Z_{\mathrm{free}}=\bar{n}\times\left(\frac{4\pi}{3}R_\mathrm{ws}^3\right).
\end{equation}
The value of the chemical potential $\mu^*$ required by the Ziman formula is related to the value of $Z^*$ by Eq.~(\ref{mu*}).\\
In most situations, in particular when the continuum of energies is uniform electron gas like, the two values are close and the impact of their difference on the Ziman resistivity is limited, due to compensation by the chemical potential \cite{Wetta2022}. In the case of argon in the thermodynamic conditions investigated in the present work, we found that $Z_\mathrm{cont}\approx Z_\mathrm{free}$ within a few percents (see Fig.~\ref{fig10}). We also did not find ``quasi-bound'' or ``quasi-free'' states which would justify corrections of $Z_\mathrm{cont}$ as recommended in our previous work on low density metallic plasmas where such states occur \cite{Wetta2023}. All our resistivity calculations were therefore performed with $Z^*=Z_\mathrm{cont}$. \\

Electrical resistivity calculations within the Ziman approach are sensitive to the ion-ion structure factor $S(q)$ mainly in the WDM conditions. In a study on aluminum at solid density and temperatures ranging from ambient one up to 100 eV, we showed the importance of equivalent modeling of $S(q)$ in the liquid and the solid states \cite{Wetta2020}. Sophisticated models are of less importance in hot plasmas, as well as in low density ones \cite{Wetta2022}. In the present work on argon, in which electronic densities remain very small (ranging from $10^{15}$ up to $10^{19}$ cm$^{-3}$), we solved the Ornstein-Zernike equation together with the hypernetted-chain closure relation for a system of screened charged spheres \cite{Rogers1980}.

\section{Shock loading experiments in presence of a magnetic field}\label{sec3}

\subsection{Description of Shilkin \emph{et al.}'s experiments \cite{Shilkin2003}}\label{subsec31}

To simplify, the experimental device consists in an approximately 30 cm long cylinder with an inner diameter of 5 cm, in which an explosive charge is placed in the first 12 up to 15 cm, and the studied gas in the remaining space. Shock waves are formed by the expansion of the detonation products in the gas. It has been checked, by a series of separate experiments, that the shock is one-dimensional and stationary at a distance of 5 up to 10 cm from the end of the charge, which allows for a uniform plasma bunch of several centimeters thick, sufficient for placing probe diagnostics. An obstacle closes the cylinder, enabling shock reflection, and further compression and heating of the studied gas. A solenoid is also wounded around the cylinder, generating a magnetic field $B=5$ T aligned along the cylinder axis.\\
Diagnostic probes provide experimental shock velocities $D$, electric resistance to an external electric current, and Hall voltage induced by the applied magnetic field. Hall voltage $\mathcal{U}_\mathrm{Hall}$ is related to the Hall coefficient $R_\mathrm{Hall}$ by
\begin{equation}
    R_\mathrm{Hall}=\dfrac{h}{Q} \dfrac{\mathcal{U}_\mathrm{Hall}}{I B},
\end{equation}
where $I$ denotes the external electric current, $h$ the plasma thickness and $Q$ a geometric factor specific to the experimental device and determined in a series of separate experiments. The Hall coefficient varies inversely to the electron density. The latter may therefore be inferred from the Hall voltage measurements.

\subsection{Rankine-Hugoniot relations in absence of a magnetic field}

\subsubsection{Incident shock wave}\label{subsec32}

\begin{figure*}[ht!]
\centering
\includegraphics[scale=0.8]{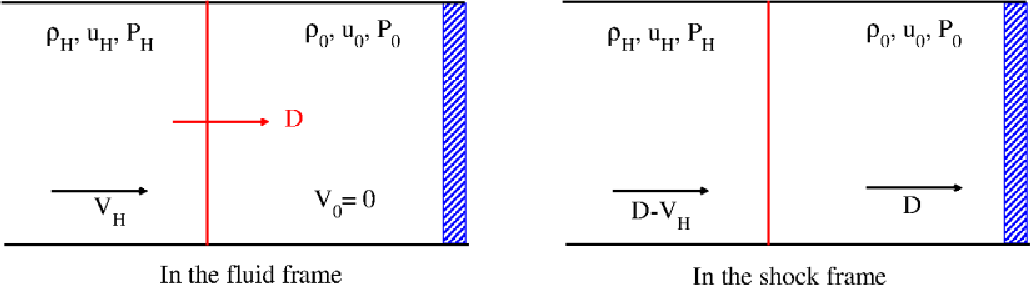}
\caption{\label{schema1} First shock wave, before reflection on the obstacle, represented in blue. Left: description in the fluid frame, where the upstream fluid velocity is $V_0=0$. Right: description in the shock frame, where the shock velocity is zero. }
\end{figure*}

In the following, the upstream (\emph{i.e.}, before the shock front) thermodynamic conditions are denoted $\rho_0$, $T_0$, $u_0$ and $P_0$, respectively being mass density, temperature, mass internal energy and pressure. $V_0$ corresponds to upstream mass velocity, and $D$ to shock velocity. Downstream (\emph{i.e.}, beyond the shock wave) thermodynamic conditions are $\rho_H$, $T_H$, $u_H$ and $P_H$, and the mass velocity $V_H$ (see Fig.~\ref{schema1}). The Rankine-Hugoniot jump conditions across a planar shock front consist of three conservation relations. In the reference frame fixed to the shock:
\begin{equation}\label{RH1}
    \rho_H(D-V_H)=\rho_0(D-V_0), 
\end{equation}
is the first relation, expressing the mass conservation across the shock front,
\begin{equation}\label{RH2}
    \rho_H(D-V_H)^2+P_H=\rho_0(D-V_0)^2+P_0, 
\end{equation}
is the momentum conservation law, and
\begin{equation}\label{RH3}
    \dfrac{1}{2}(D-V_H)^2+u_H+\dfrac{P_H}{\rho_H}= \dfrac{1}{2}(D-V_0)^2+u_0+\dfrac{P_0}{\rho_0},
\end{equation}
the energy conservation equation. Inserting Eq. (\ref{RH1}) into Eq. (\ref{RH2}) yields
\begin{equation}\label{RH4}
    (D-V_0)=\left[\dfrac{\rho_H}{\rho_0}\dfrac{(P_H-P_0)}{(\rho_H-\rho_0)}\right]^{1/2}.
\end{equation}
Using this in Eq. (\ref{RH1}) gives
\begin{equation}
    (V_H-V_0)=(D-V_0)\left(1-\dfrac{\rho_0}{\rho_H} \right),
\end{equation}
and putting Eq.(\ref{RH4}) in Eq.(\ref{RH3})
\begin{equation}
    (u_H-u_0)+\dfrac{1}{2}(P_H+P_0)\left(\dfrac{1}{\rho_H}-\dfrac{1}{\rho_0} \right)=0.
\end{equation}
The latter equation, together with an equation-of-state $u(\rho,T)$ and $P(\rho,T)$, and under the condition that Eq.(\ref{RH4}) is satisfied, yields the thermodynamic conditions $\rho_H$, $T_H$, $u_H$ and $P_H$ beyond the shock wave, and the value of the downstream velocity $V_H$ as functions of the shock velocity $D$.

\subsubsection{Reflected shock wave}\label{subsec33}

\begin{figure*}[ht!]
\centering
\includegraphics[scale=0.8]{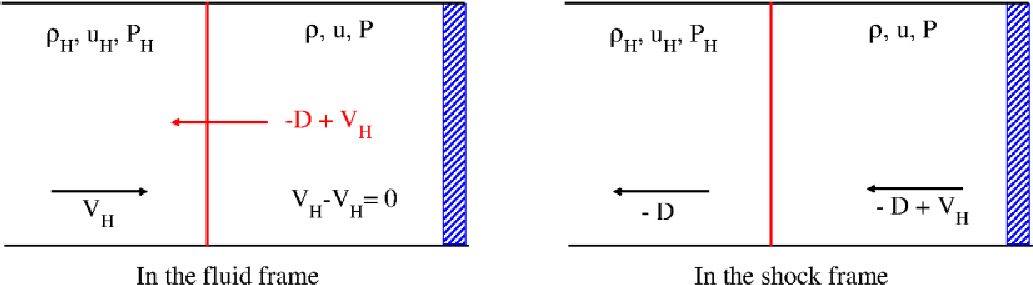}
\caption{\label{schema2} Second shock wave, after reflection on the obstacle, represented in blue. Left: description in the fluid frame, where the upstream fluid velocity is $V_0=-V_H$ and the shock velocity $-D+V_H$, $V_H$ being the fluid velocity downstream the incident shock. Right: description in the shock frame, where the shock velocity is zero.}
\end{figure*}
 
In Shilkin \emph{et al.}'s experiments, the shock wave encounters an obstacle, creating a reflected shock wave opposing to the shocked plasma moving at velocity $V_H$ (see Fig.~\ref{schema2}). The upstream conditions correspond here to the preceding downstream one, \emph{i.e.}, we have $\rho_0=\rho_H$, $T_0=T_H$, $u_0=u_H$, $P_0=P_H$ and $V_0=V_H$. In the fluid frame, the reflected shock velocity is (-$D+V_H$) (Fig.~\ref{schema2} left). We will here denote the downstream quantities $\rho$, $T$, $u$, $P$ and $V$. Assuming that the incident shock wave reflects perfectly on the obstacle, the downstream velocity is $V=-V_H+V_H=0$. Rankine-Hugoniot relations read then, in the shock frame (Fig.~\ref{schema2} right)
\begin{equation}
    \rho(D-V_H)=\rho_H D, 
\end{equation}
\begin{equation}
    \rho(D-V_H)^2+P=\rho_H D^2+P_H, 
\end{equation}
and
\begin{equation}
    \dfrac{1}{2}(D-V_H)^2+u+\dfrac{P}{\rho}= \dfrac{1}{2} D^2+u_H+\dfrac{P_H}{\rho_H}.
\end{equation}
After performing the same operations as for the incident shock wave case, we get
\begin{equation}
    (u-u_H)+\dfrac{1}{2}(P+P_H)\left(\dfrac{1}{\rho}-\dfrac{1}{\rho_H} \right)=0,
\end{equation}
which, together with our EOS $u(\rho,T)$ and $P(\rho,T)$, and under the constraint that
\begin{equation}
    V_H=\left(1-\dfrac{\rho_H}{\rho} \right)\left[\dfrac{\rho}{\rho_H}\dfrac{(P-P_H)}{(\rho-\rho_H)}\right]^{1/2},
\end{equation}
yields the thermodynamic conditions $\rho$, $T$, $u$ and $P$ beyond the reflected shock wave, as functions of the upstream mass velocity $V_H$ reached downstream the incident shock.

\subsection{Impact of the external magnetic field on Rankine-Hugoniot relations}\label{subsec34}

Up to now, we did not take into account the existence of the magnetic field present in Shilkin \emph{et al.}'s experiments for the needs for Hall effect measurements. The magnetic field interacts strongly with the plasma flow, and the Rankine-Hugoniot relations must be revised within the framework of magneto-hydrodynamic theory.\\

The flow of a compressible, non-viscous heat-insulating fluid in a magnetic field is described by a set of continuity relations \cite{Landau1987}, commonly referred to Euler's equations, coupled to Maxwell's relations. The first Euler equation expresses the conservation of mass:
\begin{equation}\label{Euler1}
    \dfrac{\partial\rho}{\partial t}+\Vec{\nabla}\cdot (\rho\Vec{V})=0.
\end{equation}
The second relation is the actual Euler's equation \cite{Euler1757}, and expresses the conservation of momentum:
\begin{equation}\label{Euler2}
    \dfrac{\partial}{\partial t}(\rho\Vec{V})+\Vec{\nabla}\cdot(\rho\Vec{V}\otimes\Vec{V})=-\Vec{\nabla}P+\Vec{S},
\end{equation}
where $\Vec{S}$ denotes the external forces acting on the fluid. Alternately, this conservation relation may also be written:
\begin{equation}\label{Euler2b}
    \dfrac{\partial}{\partial t}(\rho\Vec{V})+(\rho\Vec{V}\cdot\Vec{\nabla})\Vec{V}=-\Vec{\nabla}P+\Vec{S}.
\end{equation}
$\Vec{S}$ is here the sum of the Lorentz forces acting on both electrons and ions:
\begin{align}
    \Vec{S} &= \Vec{F_i}+\Vec{F_e}\\
    &=Z^* e n_i (\Vec{E}+\Vec{V_i}\times\Vec{B}) - e n_e (\Vec{E}+\Vec{V_e}\times\Vec{B}) \nonumber\\
    &= en_e (\Vec{V_i}-\Vec{V_e})\times\Vec{B}, \nonumber
\end{align}
where the last equality results from the electrical neutrality assumption $Z^* n_i=n_e$. Finally, introducing the current density $\Vec{J}=en_e (\Vec{V_i}-\Vec{V_e})$:
\begin{equation}
    \Vec{S}=\Vec{J}\times\Vec{B}.
\end{equation}
The energy conservation relation follows from the derivation of two equations expressing the balance of the kinetic energy and the internal energy components. The kinetic energy conservation equation reads \cite{Landau1987}
%
\begin{equation}\label{Euler3}
    \dfrac{\partial}{\partial t}\left(\dfrac{1}{2}\rho V^2\right)+\Vec{\nabla}\cdot\left[\left(\dfrac{1}{2}\rho V^2\right)\Vec{V}\right]=-\Vec{\nabla}P\cdot\Vec{V}+W,
\end{equation}
$W$ being the work produced by the Lorentz forces:
\begin{align}\label{W}
    W &= \Vec{F_i}\cdot\Vec{V_i}+\Vec{F_e}\cdot\Vec{V_e}\nonumber\\
      &= \Vec{E}\cdot\Vec{J}.
\end{align}
The derivation of a similar conservation relation for the internal energy \cite{Landau1987} starts by writing the first law of thermodynamics:
\begin{equation}
    dU=-Pd\Omega,
\end{equation}
where $U$ denotes the internal energy and $\Omega$ the volume. Introducing the mass internal energy $u=U/m_e$:
\begin{equation}
    \dfrac{du}{dt}=\dfrac{P}{\rho^2}\dfrac{d\rho}{dt}.
\end{equation}
Applying the relation
\begin{equation}
    \dfrac{d}{dt}=\dfrac{\partial}{\partial t}+\Vec{V}\cdot\Vec{\nabla}
\end{equation}
to both sides of this equation, and using the mass conservation equation (\ref{Euler1}) yields:
\begin{equation}
   \dfrac{\partial u}{\partial t} +\Vec{V}\cdot\Vec{\nabla}u=-\dfrac{P}{\rho}\Vec{\nabla}\cdot\Vec{V}.
\end{equation}
The internal energy balance equation is obtained by summing the preceding equation multiplied by $\rho$ and Eq.~(\ref{Euler1}) multiplied by $u$:
\begin{equation}\label{Euler3_2}
    \dfrac{\partial}{\partial t}(\rho u)+\Vec{\nabla}\cdot(u\rho\Vec{V})=-P\Vec{\nabla}\cdot\Vec{V},
\end{equation}
Finally, one gets, adding Eqs.~(\ref{Euler3}) (also using Eq.~(\ref{W})) and (\ref{Euler3_2}):
\begin{equation}\label{Euler3_3}
    \dfrac{\partial}{\partial t} \left( \dfrac{1}{2}\rho V^2+\rho u\right) +\Vec{\nabla}\cdot\left[\left(\dfrac{1}{2}\rho V^2+\rho h)\Vec{V} \right)\right]=\Vec{E}\cdot\Vec{J},
\end{equation}
$h$ denoting the mass enthalpy $h=u+\dfrac{P}{\rho}$.\\

The Maxwell equations read
\begin{gather}
    \dfrac{\partial\Vec{B}}{\partial t}=-\Vec{\nabla}\times\Vec{E},\label{Maxwell1}\\
    \Vec{\nabla}\times\Vec{B}=\mu_0 \Vec{J},\label{Maxwell2}\\
    \Vec{\nabla}\cdot\Vec{B}=0,\nonumber\\
    \Vec{\nabla}\cdot\Vec{E}=0,\nonumber 
\end{gather}
$\mu_0=4\pi 10^{-7}$ H/m denoting the magnetic permeability. The two first Eqs.~(\ref{Maxwell1}) and (\ref{Maxwell2}) link electrical field $\Vec{E}$, magnetic induction $\Vec{B}$ and electric current density $\Vec{J}$. The Ohm's law provides the third relation necessary to obtain these quantities. Assuming that Hall effects can be neglected (this point will be verified later), Ohm's law reads
\begin{equation}\label{Ohm}
    \Vec{J}=\sigma (\Vec{E}+\Vec{V}\times\Vec{B}).
\end{equation}
Using Eq.~(\ref{Maxwell2}):
\begin{equation}
    \Vec{J}\times\Vec{B}=\dfrac{\Vec{\nabla}\times\Vec{B}}{\mu_0}\times\Vec{B}=-\Vec{\nabla}\left(\dfrac{B^2}{2\mu_0}\right)+\dfrac{(\Vec{B}\cdot\Vec{\nabla})\Vec{B}}{\mu_0}.
\end{equation}
The quantity on the right side of this equation is the divergence $\Vec{\nabla}\cdot\overline{\overline{T}}_\mathrm{em}$ of the magnetic pressure tensor (also called the Maxwell tensor):
\begin{equation}
    \overline{\overline{T}}_\mathrm{em}=-\dfrac{B^2}{2\mu_0}\overline{\overline{I}}+\dfrac{\Vec{B}\otimes\Vec{B}}{\mu_0}, 
\end{equation}
whose elements read
\begin{equation}
    T_{ij}=-\dfrac{B^2}{2\mu_0}\delta_{ij}+\dfrac{B_i B_j}{\mu_0}.
\end{equation}
The momentum conservation equation Eq.~(\ref{Euler2}) reads then
\begin{align}
    \dfrac{\partial}{\partial t}(\rho\Vec{V})+&\Vec{\nabla}\cdot(\rho\Vec{V}\otimes\Vec{V})+\Vec{\nabla}P\nonumber\\
    &=-\Vec{\nabla}\left(\dfrac{B^2}{2\mu_0} \right)+\dfrac{\Vec{\nabla}\cdot(\Vec{B}\otimes\Vec{B})}{\mu_0},
\end{align}
or, alternately:
\begin{align}\label{Euler2_2}
    \dfrac{\partial}{\partial t}(\rho\Vec{V})+&(\rho\Vec{V}\cdot\Vec{\nabla})\Vec{V}+\Vec{\nabla}P\nonumber\\
    &=-\Vec{\nabla}\left(\dfrac{B^2}{2\mu_0} \right)+\dfrac{(\Vec{B}\cdot\Vec{\nabla})\Vec{B}}{\mu_0}.
\end{align}
Let us here gather the equations that will be useful for our further developments:
\begin{gather}
    \dfrac{\partial \rho}{\partial t}+\Vec{\nabla}\cdot(\rho\Vec{V})=0,\\
    \dfrac{\partial}{\partial t}(\rho\Vec{V})+(\rho\Vec{V}\cdot\Vec{\nabla})\Vec{V}+\Vec{\nabla}P=-\Vec{\nabla}\left(\dfrac{B^2}{2\mu_0} \right)+\dfrac{(\Vec{B}\cdot\Vec{\nabla})\Vec{B}}{\mu_0},\\
    \dfrac{\partial}{\partial t} \left( \dfrac{1}{2}\rho V^2+\rho u\right) +\Vec{\nabla}\cdot\left[\left(\dfrac{1}{2}\rho V^2+\rho h)\Vec{V} \right)\right]=\Vec{E}\cdot\Vec{J},\\
    \Vec{J}=\sigma (\Vec{E}+\Vec{V}\times\Vec{B}).
\end{gather}
We will now consider the propagation of a planar shock in a cylindrical shock tube surrounded by a solenoid. The latter induces a magnetic field $\Vec{B_0}$ parallel to the cylindrical axis, which we will identify to the $z$ axis. The upstream conditions will be denoted $\rho_1$, $P_1$ and $h_1$ as concerns the density, the pressure and the enthalpy, the gas velocity and the magnetic field being respectively: $\Vec{V_1}=(0,0,V_1)$ and $\Vec{B_1}=(0,0,B_0)$. Their downstream counterparts will be: $\rho_2,\,P_2,\, h_2,\, \Vec{V_2}=(0,0,V_2)$ and $\Vec{B_2}=(0,0,B_2)$. We will also assume low upstream electrical conductivity, \emph{i.e.}, $\sigma_1\approx 0$, and the possibility of a strong increase of this quantity inside the shock front. Therefore, inside the shock, the gas velocity and the magnetic field get components parallel to the shock plane, and read: $\Vec{V}=(V_{||},0,V_\perp)$ and $\Vec{B}=(B_{||},0,B_\perp)$, where the subscripts $\perp$ and $||$ respectively denote the components perpendicular and parallel to the shock plane. The increase of the electrical conductivity induces an electric current $\Vec{J}$ in the shock plane, which in turn induces the parallel to the shock plane contribution to the magnetic field. The relation between $B_{||}$ and $\Vec{J}$ results from Maxwell-Amp\`ere equation (\ref{Maxwell2}) and reads
\begin{equation}
    \Vec{J}=\left(0,\dfrac{1}{\mu_0}\dfrac{\partial B_{||}}{\partial z},0 \right).
\end{equation}
Finally, assuming that the shock plane remains normal to the $z$ direction, all quantities involved in the problem are supposed to vary only with $z$ coordinate, \emph{i.e}: $\dfrac{\partial}{\partial x}\equiv 0$ and $\dfrac{\partial}{\partial y}\equiv 0$. We also consider stationary shock, \emph{i.e}: $\dfrac{\partial}{\partial t}\equiv 0$.\\
Introducing the notation $[\mathcal{F}]_{z_0}^{z_0+\delta}=\mathcal{F}(z_0+\delta)-\mathcal{F}(z_0)$, $z_0$ denoting the position of the upstream front and $\delta$ the shock thickness:
\begin{gather}
    [\rho V]_{z_0}^{z_0+\delta}=0,\\
    \left[\rho V^2+P\right]_{z_0}^{z_0+\delta}=\int_{z_0}^{z_0+\delta}\left[-\dfrac{\partial}{\partial z}\left(\dfrac{B^2}{2\mu_0}\right) +\dfrac{B_\perp}{\mu_0}\dfrac{\partial B_\perp}{\partial z}\right]\,dz,\\
    \left[\left(\dfrac{1}{2}\rho V^2+h \right)V\right]_{z_0}^{z_0+\delta}=\int_{z_0}^{z_0+\delta} \left(\dfrac{1}{\mu_0}\dfrac{\partial B_{||}}{\partial z} \right) \left[\dfrac{1}{\sigma\mu_0}\dfrac{\partial B_{||}}{\partial z}-(V_{||} B_\perp-V_\perp B_{||})\right]\,dz.
\end{gather}
Maxwell relation $\Vec{\nabla}\cdot\Vec{B}=0$ yields $\dfrac{\partial B_\perp}{\partial z}=0$, and the second of the above equations becomes:
\begin{equation}
    \left[\rho V^2+P+\left(\dfrac{B_{||}^2}{2\mu_0}\right)\right]_{z_0}^{z_0+\delta}=0.
\end{equation}
When the electrical conductivity $\sigma$ inside the shock remains low, the above equations reduce to the usual Rankine-Hugoniot jump relations (\ref{RH1}), (\ref{RH2}) and (\ref{RH3}), applicable to shock propagation in absence of any magnetic field. Indeed, in that situation, no electric current can appear, and consequently: $\dfrac{\partial B_{||}}{\partial z}=0$. Shilkin \emph{et al.} assert that this is the case for their direct shock experiments on argon.\\ 
At the opposite, they claim that all of their reflected shocks are ionizing, and that electrical conductivities inside these shocks are then high enough to reach the conditions for ``frozen-in'' magnetic fields. Let us consider the extreme case of infinite electrical conductivity $\sigma$. We have then:
\begin{equation}
    \dfrac{1}{\sigma\mu_0}\dfrac{\partial B_{||}}{\partial z} \ll -(V_{||}B_\perp-V_\perp B_{||}),
\end{equation}
and also $B_{||}\gg B_\perp$, yielding: $V_{||}B_\perp-V_\perp B_{||}\approx -V_\perp B_{||}$. The electric field inside the shock is $\Vec{E}=-\Vec{V}\times\Vec{B}\approx (0,-V_\perp B_{||},0)$. Applying Maxwell relation $\Vec{\nabla}\times\Vec{E}=0$ one obtains that $V_\perp B_{||}$ is constant across the shock front. The jump relations in the limit $\sigma\rightarrow\infty$ read then
\begin{gather}
    [\rho V]_{z_0}^{z_0+\delta}=0,\\
    [B_{||}V]_{z_0}^{z_0+\delta}=0,\\
    \left[\rho V^2+P+\left(\dfrac{B_{||}^2}{2\mu_0}\right)\right]_{z_0}^{z_0+\delta}=0,\\
    \left[\left(\dfrac{1}{2}\rho V^2+h +\dfrac{B_{||}^2}{\mu_0}\right)V\right]_{z_0}^{z_0+\delta}=0.
\end{gather}
Since the gas velocities $\Vec{V}$ are taken at their upstream and downstream values (that are parallel to the $z$ axis), $V_\perp$ has been replaced by $V$ in these equations. The two first relations can also be gathered to give:
\begin{equation}
    \left[\dfrac{B_{||}}{\rho}\right]_{z_0}^{z_0+\delta}=0. 
\end{equation}
The equality $[\rho V]_{z_0}^{z_0+\delta}=[B_{||} V]_{z_0}^{z_0+\delta}$ is at the origin of the expression ``frozen-in'' magnetic field, since the evolution of $B_{||}$ follows exactly the one of the mass, as if magnetic lines are attached to matter.\\

$\sigma$ is the key quantity that governs the evolution of the magnetic field inside the shock front. Using Eqs.~(\ref{Maxwell1}), (\ref{Ohm}) and (\ref{Maxwell2}) one gets the following induction equation:
\begin{align}\label{diff_eq}
    \dfrac{\partial \Vec{B}}{\partial t}&=-\Vec{\nabla}\times\Vec{E}\nonumber\\
    &=-\Vec{\nabla}\times\left(\dfrac{\Vec{J}}{\sigma}-\Vec{V}\times\Vec{B}\right)\\
    &=-\dfrac{\Vec{\nabla}\times(\Vec{\nabla}\times\Vec{B})}{\mu_0\sigma}+\Vec{\nabla}\times(\Vec{V}\times\Vec{B}).\nonumber
\end{align}
The first term in the right member of the latter equation describes diffusion of the magnetic field, and the second term its advection by the fluid's motion.\\
The magnetic Reynolds number, defined as the ratio:
\begin{equation}
    \mathcal{R}_m\equiv\dfrac{|\Vec{\nabla}\times(\Vec{V}\times\Vec{B})|}{|\eta\Vec{\nabla}^2\Vec{B}|}
\end{equation}
measures the relative importance of advection and diffusion of the magnetic field. The parameter $\eta=\dfrac{1}{\mu_0\sigma}$ (units: m$^2$/s) is the magnetic diffusivity. $\mathcal{R}_m$ is of the order of 
\begin{equation}
    \mathcal{R}_m\approx\dfrac{VB/L_\mathrm{adv}}{\eta/L_\mathrm{dif}^2},
\end{equation}
where $L_\mathrm{adv}$ and $L_\mathrm{dif}$ are characteristic length scales for respectively advection and diffusion phenomena. In the absence of shock wave perturbing the magnetized fluid, the two lengths may be considered equal, and the Reynolds number is given by the formula $\mathcal{R}_m=\mu_0\sigma V L$.\\
In presence of a shock wave, the typical advection length is the thickness $\delta$ of the shock front, while the characteristic diffusion length is the dimension of the material through which the magnetic field passes, in the case of Shilkin's experiments the diameter $d$ of the cylinder containing the argon gas. We therefore write the Reynolds number in the context of shock experiments as:
\begin{equation}
    \mathcal{R}_m^*=\dfrac{VB/\delta}{\eta/d^2}=\mu_0\sigma V \dfrac{d^2}{\delta}.
\end{equation}
From Shilkin \emph{et al.}'s paper \cite{Shilkin2003}, we estimate typical velocity $V\approx$ 2.5 km/s, and note experimental electrical conductivities $1\,(\Omega \text{m})^{-1}\lesssim\sigma\lesssim 10^3\,(\Omega \text{m})^{-1}$ downstream incident shock wave and $10^3\,(\Omega \text{m})^{-1}\lesssim\sigma\lesssim 10^4\,(\Omega \text{m})^{-1}$ behind the reflected shock wave. Experimentally, the shock thickness for argon at Mach numbers lying between 2 and 11, is in the range 3.7 mm $\lesssim \delta \lesssim$ 5.5 mm \cite{Alsmeyer1976,Robben1966,Schmidt1969}. Retaining $\delta=5$ mm, and the diameter $d=5$ cm of the cylindrical experimental device containing the argon gas, we get $0.16\times 10^{-2}\lesssim\mathcal{R}_m^*\lesssim 0.16$ downstream the incident shock wave, and $0.16\lesssim\mathcal{R}_m^*\lesssim 16$ behind the reflected one. These latter values are consistent with Shilkin \emph{et al.}'s assertion that the magnetic field is ``frozen-in'' in their reflected shock experiments. We therefore introduce the magnetic pressures and mass internal energies in the Rankine-Hugoniot relations for the reflected shock. Noting respectively $B$ and $B_0$ the magnetic induction downstream and upstream the reflected shock wave, they read \cite{DeHoffmann1950,Helfer1953,Kennel1994,Mallick2011,Berton2021} 
\begin{equation}
    \rho(D-V_H)=\rho_H D, 
\end{equation}
\begin{equation}
    \rho(D-V_H)^2+P+\dfrac{B^2}{2\mu_0}=\rho_H D^2+\left(P_H+\dfrac{B_0^2}{2\mu_0}\right), 
\end{equation}
and
\begin{equation}
    \dfrac{1}{2}(D-V_H)^2+u+\dfrac{P}{\rho}+\dfrac{B^2}{\mu_0\rho}= \dfrac{1}{2} D^2+u_H+\dfrac{P_H}{\rho_H}+\dfrac{B_0^2}{\mu_0\rho_H}.
\end{equation}
The jump relation relative to the magnetic fields reads
\begin{equation}\label{ideal_MHD}
    \rho_0 B_H=\rho_H B_0.
\end{equation}

\subsection{Derivation of the resistivity tensor from the Boltzmann theory}\label{subsec41}

In presence of a magnetic field, Ohm's law is no longer linear. The electric current and electric field are then related by the conductivity tensor $\bar{\bar{\sigma}}$:
\begin{equation}
    \Vec{J}=\bar{\bar{\sigma}}\Vec{E}\ \Leftrightarrow \,\Vec{E}=\bar{\bar{\eta}}\Vec{J},
\end{equation}
where we have introduced the resistivity tensor $\bar{\bar{\eta}}$, \emph{i.e.}, the inverse of the conductivity tensor. The conductivity tensor derives from the Boltzmann transport theory, in terms of powers of the collision times. This will be the object of the first subsection. In the second one, we will discuss the possibility of calculating effective collisions times in the Ziman-Evans average-atom approach used in our work.\\

The Boltzmann equation for electron transport reads \cite{Ebeling1984}:
\begin{equation}
    \dfrac{\partial f_e}{\partial t}+\Vec{v}\cdot\Vec{\nabla}f_e-e\left(\Vec{E}+\Vec{v}\times\Vec{B}\right)\dfrac{\partial f_e}{\partial \Vec{p}}=I[f_e],
\end{equation}
where $f_e$ is the electron distribution function, $\Vec{v}$ the electron velocity, $\Vec{p}=m_e\Vec{v}$ the electron momentum, and $I[f_e]$ the collision integral. In the context of electric conduction it is assumed that $f_e$ is independent of time, and that it varies in space only through a possible temperature gradient, \textit{i.e.}, that
\begin{equation}
    \dfrac{\partial f_e}{\partial\Vec{r}}=\dfrac{\partial f}{\partial T}\dfrac{\partial T}{\partial\Vec{r}}.
\end{equation}
In absence of such a gradient, the two first terms of the left-hand side of the Boltzmann equation can then be dropped. 
Another simplification consists in considering small perturbations around the Fermi-Dirac distribution function Eq.~(\ref{fd}): $f_e=f(\epsilon,\mu)+\delta f$, with
\begin{equation}
    \delta f=-\phi\dfrac{\partial f(\epsilon,\mu)}{\partial\epsilon},
\end{equation}
introducing a quantity $\phi$ depending on the configuration variables.
Using the equality:
\begin{equation}
    \dfrac{\partial f}{\partial \Vec{p}}=\Vec{v}\,\dfrac{\partial f}{\partial\epsilon}
\end{equation}
and only retaining the first order in $\phi$, the Boltzmann equation reduces then to the following linearized form: 
\begin{equation}
    e\Vec{v}\cdot\Vec{E}\dfrac{\partial f}{\partial\epsilon}-e[\Vec{v}\times\Vec{B}]\dfrac{\partial f}{\partial\epsilon}\dfrac{\partial \phi}{\partial \Vec{p}}=-\mathcal{L}[I],
\end{equation}
$\mathcal{L}[I]$ denoting the linearized collision integral. Further writing \cite{Harutyunyan2016}: 
\begin{equation}
    \phi={\Vec{p}}.{\Vec{\xi}}(\epsilon),
\end{equation}
the linearized collision term takes the following form, in terms of the collision time $\tau(\epsilon)$:
\begin{equation}
    \mathcal{L}[I]={\Vec{\xi}}.{\Vec{p}}\dfrac{1}{\tau(\epsilon)}.
\end{equation}
The linearized electron transport equation then reads \cite{Harutyunyan2016}
\begin{equation}\label{ttau}
    e{\Vec{v}}.[{\Vec{E}}+({\Vec{\xi}}\times{\Vec{B}})]=-{\Vec{\xi}}.{\Vec{p}}\dfrac{1}{\tau(\epsilon)}.
\end{equation}
The most general decomposition of ${\Vec{\xi}}$ is as follows:
\begin{equation}
    {\Vec{\xi}}=\alpha {\hat{e}}+\zeta {\hat{b}}+\gamma[{\hat{e}}\times{\hat{b}}],
\end{equation}
with ${\hat{b}}={\Vec{B}}/B$ and ${\hat{e}}={\Vec{E}}/E$. 
Using this form in Eq.~(\ref{ttau}), one gets (we have dropped the energy dependence of $\tau(\epsilon)$ to lighten the formulas):
\begin{equation}
    \alpha=-eE\dfrac{\tau}{m_e(1+\omega_c^2\tau^2)},
\end{equation}
\begin{equation}
    \dfrac{\zeta}{\alpha}=(\omega_c\tau)^2({\hat{e}}.{\hat{b}})
\end{equation}
and 
\begin{equation}
    \dfrac{\gamma}{\alpha}=-\omega_c\tau,
\end{equation}
where $\omega_c=eB/m_e$ is the cyclotron frequency, and finally \cite{Harutyunyan2016}:
\begin{equation}\label{phi}
    \phi=-\dfrac{e\tau}{1+(\omega_c\tau)^2}v_i\left[\delta_{ij}-\omega_c\tau\epsilon_{ijk}b_k+(\omega_c\tau)^2b_ib_j\right]E_j.
\end{equation}
The symbol $\varepsilon_{ijk}$ represents the usual Levi-Civita one \cite{Tyldesley1973}:
\begin{equation}
    \varepsilon_{ijk}={\begin{cases}+1&{\text{if }}(i,j,k){\text{ is }}(1,2,3),(2,3,1),{\text{ or }}(3,1,2),\\-1&{\text{if }}(i,j,k){\text{ is }}(3,2,1),(1,3,2),{\text{ or }}(2,1,3),\\\;\;\,0&{\text{if }}i=j,{\text{ or }}j=k,{\text{ or }}k=i.\end{cases}}
\end{equation}
That is, $\varepsilon_{ijk}$ is equal to 1 if $(i, j, k)$ is an even permutation of $(1, 2, 3)$, to -1 if it is an odd permutation, and to 0 if any index is repeated. The cyclic permutations of $(1, 2, 3)$ are all even permutations, similarly the anti-cyclic permutations are all odd permutations. 
The derivation of the electric current vector $\Vec{J}$ follows immediately. Its components are:
\begin{align}
    J_i&=\dfrac{2}{(2\pi)^3}\int e v_i \delta f d^3k\\
    &=\dfrac{2}{(2\pi)^3}\int e v_i \phi \dfrac{\partial f}{\partial\epsilon} d^3k.\nonumber
\end{align}
Using Eq.~(\ref{phi}):
\begin{equation}
    J_i=\sigma_{ij}E_j,
\end{equation}
where $\sigma_{ij}$ are the component of the conductivity tensor, that read
\begin{equation}
    \sigma_{ij}=\delta_{ij}\sigma_0-\epsilon_{ijm}b_m\sigma_1+b_ib_j\sigma_2.
\end{equation}
When the magnetic field is along the $z$ direction:
\begin{equation}
    \Bar{\Bar{\sigma}}=\left(
    \begin{array}{ccc}
    \sigma_0 & -\sigma_1 & 0\\
    \sigma_1 & \sigma_0 & 0\\
    0 & 0 & \sigma_0+\sigma_2\\
    \end{array}
    \right)
\end{equation}
where
\begin{equation}
    \sigma_n=\dfrac{4e^2}{3h^3m_e}\int_0^{\infty} \dfrac{p^2}{2m_e}\dfrac{\tau(\omega_c\tau)^n}{1+(\omega_c\tau)^2}\left(-\dfrac{\partial f}{\partial \epsilon} \right)4\pi p^2dp
\end{equation}
or
\begin{equation}
    \sigma_n=\dfrac{e^2}{3\pi^2m_e}\int_0^{\infty} k^3\dfrac{\tau(\omega_c\tau)^n}{1+(\omega_c\tau)^2}\left(-\dfrac{\partial f}{\partial \epsilon} \right)d\epsilon.
\end{equation}
The latter form can also be rewritten as
\begin{equation}
 \sigma_n=\dfrac{e^2 n_e}{m_e} \Big\langle \dfrac{\tau(\omega_c\tau)^n}{1+(\omega_c\tau)^2}\Big\rangle, 
\end{equation}
where
\begin{equation}
 \Big\langle \dfrac{\tau(\omega_c\tau)^n}{1+(\omega_c\tau)^2}\Big\rangle=\dfrac{1}{3\pi^2 n_e} \int_0^{\infty} k^3\dfrac{\tau(\omega_c\tau)^n}{1+(\omega_c\tau)^2}\left(-\dfrac{\partial f}{\partial \epsilon} \right)d\epsilon.
\end{equation}
The resistivity tensor is the inverse of the conductivity one:
\begin{equation}\renewcommand{\arraystretch}{2}
    \Bar{\Bar{\eta}}=\left(
    \begin{array}{ccc}
    \displaystyle\dfrac{\sigma_0}{\sigma_0^2+\sigma_1^2} & \displaystyle\dfrac{\sigma_1}{\sigma_0^2+\sigma_1^2} & 0\\
    - \displaystyle\dfrac{\sigma_1}{\sigma_0^2+\sigma_1^2}& \displaystyle\dfrac{\sigma_0}{\sigma_0^2+\sigma_1^2} & 0\\
    0 & 0 & \dfrac{1}{\sigma_0+\sigma_2}\\
    \end{array}
    \right).
\end{equation}

\subsection{Resistivity tensor at the limit $\omega_c\tau\ll 1$}\label{subsec42}

One has, at order 2 in the expansion in $(\omega_c \tau)$:
\begin{equation}
    \sigma_0\approx \dfrac{e^2 n_e}{m_e}\left(\langle\tau\rangle-\omega_c^2\langle\tau^3\rangle \right) +O(\omega_c^3),
\end{equation}
and
\begin{equation*}
    \sigma_1\approx \dfrac{e^2 n_e}{m_e}\omega_c\langle\tau^2\rangle+O(\omega_c^3),
\end{equation*}
which yields
\begin{equation}
    \dfrac{\sigma_0}{\sigma_0^2+\sigma_1^2}\approx
    \dfrac{m_e}{e^2n_e}\dfrac{1}{\langle\tau\rangle}\ \left\{1+\omega_c^2\left(\dfrac{\langle\tau^3\rangle}{\langle\tau\rangle}-\dfrac{\langle\tau^2\rangle^2}{\langle\tau\rangle^2} \right) \right\}+O(\omega_c^3),
\end{equation}
and
\begin{align}
    \dfrac{\sigma_1}{\sigma_0^2+\sigma_1^2}
    \approx&\dfrac{m_e\omega_c}{e^2n_e}\dfrac{\langle\tau^2\rangle}{\langle\tau\rangle^2}\nonumber\\
    &\times\left\{1+\omega_c^2\left(2\dfrac{\langle\tau^3\rangle}{\langle\tau\rangle}-\dfrac{\langle\tau^2\rangle^2}{\langle\tau\rangle^2} \right) \right\}+O(\omega_c^4).
\end{align}
Noting that Eq.~(\ref{eta}) can be rewritten:
\begin{equation}
    \eta=\dfrac{m_e}{n_e e^2}\Big\langle\dfrac{v}{\Lambda(\epsilon)}\Big\rangle,
\end{equation}
with
\begin{equation}
   \Big\langle\dfrac{v}{\Lambda(\epsilon)}\Big\rangle = \dfrac{1}{3\pi^2 n_e} \int_0^\infty k^3 \dfrac{\partial f}{\partial\epsilon} \dfrac{v}{\Lambda(\epsilon)} d\epsilon,
\end{equation}
where $\Lambda(\epsilon)$ is related to $\mathcal{I}(\epsilon)$ by
\begin{equation}\label{Lambda}
    \displaystyle\dfrac{1}{\Lambda(\epsilon)}=\displaystyle\dfrac{\pi n_i}{k^4}\mathcal{I}(\epsilon),
\end{equation}
one gets a collision time $\tau(\epsilon)$ from the data provided by our average-atom code
\begin{equation}\label{tau}
    \tau(\epsilon)=\displaystyle\dfrac{\Lambda(\epsilon)}{v}.
\end{equation}
In the conditions of the experiments performed by Shilkin \emph{et al.} on argon, we calculated this way values of the order of (in atomic units) $\langle\tau\rangle\approx 5\,10^2$, $\langle\tau^2\rangle\approx 3\,10^5$ and $\langle\tau^3\rangle\approx 3\,10^8$ for the shock velocity $D$=2.5 km/s. At the experimental magnetic induction $B=5$ T, the cyclotron frequency is $\omega_c\approx 2\times 10^{-5}\text{a.u.}$, which justifies the fact that the terms in $\omega_c^2$ inside the brackets can be neglected.
Finally, since
\begin{equation}
    \dfrac{1}{\sigma_0+\sigma_2}\approx \dfrac{m_e}{e^2 n_e}\dfrac{1}{\langle\tau\rangle}+O(\omega_c^5),
\end{equation}
the resistivity tensor reads, in the conditions of Shilkin \emph{et al.} experiments
\begin{equation}\renewcommand{\arraystretch}{2}
    \Bar{\Bar{\eta}}\approx\left(
    \begin{array}{ccc}
    \displaystyle\dfrac{1}{\sigma} & R_\mathrm{Hall} B & 0\\
    - R_\mathrm{Hall} B & \displaystyle\dfrac{1}{\sigma} & 0\\
    0 & 0 & \dfrac{1}{\sigma}\\
    \end{array}
    \right)+O(\omega_c^2),
\end{equation}
where $R_\mathrm{Hall}$ denotes the Hall coefficient, given by:
\begin{equation}\label{Hall}
    R_\mathrm{Hall}=\dfrac{1}{e n_e}\dfrac{\langle\tau^2\rangle}{\langle\tau\rangle ^2}.
\end{equation}
In other words, the tension measured along the $\hat{x}$ direction in these experiments gives the value of the electrical resistivity of the plasma unperturbed by a magnetic field, which then can be used for comparisons to calculations with Ziman formula (\ref{eta}).

\subsection{Hall coefficient using the average-atom approach}\label{subsec44}

Using Eqs.~(\ref{tau}) and (\ref{Lambda}) in Eq.~(\ref{Hall}) yields
\begin{equation}
    R_\mathrm{Hall}=\dfrac{1}{e n_e}\times 3\pi^2 n_e \dfrac{\int_0^\infty k^3 \left(-\dfrac{\partial f}{\partial\epsilon}\right)\left(\dfrac{\Lambda}{v}\right)^2 d\epsilon}{\left[\int_0^\infty k^3 \left(-\dfrac{\partial f}{\partial\epsilon}\right)\left(\dfrac{\Lambda}{v}\right) d\epsilon\right]^2}.
\end{equation}
The dimensionless Hall constant $r_\mathrm{Hall}$ is obtained by multiplying $R_\mathrm{Hall}$ by $en_e$, \emph{i.e.}, $r_\mathrm{Hall}=R_\mathrm{Hall}\times en_e$.\\

At both solid state and non-degenerate plasma limits, the formula recovers the expected $r_\mathrm{Hall}=1$ value. In the former case
\begin{equation}
    \lim_{T\rightarrow 0}\left(-\dfrac{\partial f}{\partial \epsilon}\right) =\delta(\epsilon-\epsilon_F),
\end{equation}
yielding, using the equality $k_F^3=3\pi^2 n_e$:
\begin{equation}
    \lim_{T\rightarrow 0}r_\mathrm{Hall}=(3\pi^2 n_e)\dfrac{k_F^3\tau^2(\epsilon_F)}{k_F^6\tau^2(\epsilon_F)}=1.
\end{equation}
The plasma electron degeneracy parameter is defined as the ratio of thermal energy on Fermi energy
\begin{equation}
  \Theta=\dfrac{k_B T}{\epsilon_F}=\dfrac{2 m_e}{\hbar^2}\dfrac{k_B T}{(3\pi^2 n_e)^{2/3}}.  
\end{equation}
In the non-degenerate plasma limit $\Theta\gg 1$
\begin{equation}
   -\dfrac{\partial f}{\partial\epsilon} \rightarrow \beta e^{-\beta (\epsilon-\mu)}.
\end{equation}
$\beta e^{-\beta (\epsilon-\mu)} \ll 1$ and is a decreasing function of $\epsilon$. Expanding the scattering phase-shifts $\delta_\kappa(k)$ and the ion-ion structure factor $S(k)$ in powers of $k$, and only retaining the first terms 
\begin{equation}
  \delta_\kappa(k)\propto k \text{ and }  S(k)\propto k,
\end{equation}
$I(\epsilon)$ varies as $k^3$ and $\Lambda(\epsilon)\propto k$. Then $\tau(\epsilon)\approx C+O(k)$, $C$ denoting a constant, and, at the lowest order in the expansion
\begin{equation}
    \lim_{\Theta\gg 1}r_\mathrm{Hall}=(3\pi^2 n_e)\dfrac{\int_0^\infty k^3\left(-\dfrac{\partial f}{\partial\epsilon}\right)C^2 d\epsilon}{\left[\int_0^\infty k^3\left(-\dfrac{\partial f}{\partial\epsilon}\right)C d\epsilon\right]^2}=1,
\end{equation}
using the equality $\int_0^\infty k^3\left(-\dfrac{\partial f}{\partial\epsilon}\right)d\epsilon=3\pi^2 n_e$.\\

The value $r_\mathrm{Hall}=1$ is also predicted in the strong magnetic fields ($\omega_c\tau\gg 1$). Indeed, at order 4 in the expansions in $1/(\omega_c\tau)$:
\begin{equation}
\sigma_0\approx\dfrac{e^2 n_e}{m_e} \dfrac{1}{\omega_c^2}\left(\Big\langle\dfrac{1}{\tau}\Big\rangle-\dfrac{1}{\omega_c^2}\Big\langle\dfrac{1}{\tau^3} \Big\rangle \right), 
\end{equation}
\begin{equation}
\sigma_1\approx \dfrac{e^2 n_e}{m_e} \dfrac{1}{\omega_c}\left(1-\dfrac{1}{\omega_c^2}\Big\langle\dfrac{1}{\tau^2} \Big\rangle \right),   
\end{equation}
which yield, only retaining the most important term when calculating the ratio $\dfrac{\sigma_1}{\sigma_0^2+\sigma_1^2}$
\begin{equation}
\lim_{\omega_c\tau\gg 1} r_\mathrm{Hall} = 1-\dfrac{1}{\omega_c^2}\left(\Big\langle\dfrac{1}{\tau} \Big\rangle^2-\Big\langle\dfrac{1}{\tau^2} \Big\rangle \right).  
\end{equation}

\section{Electrical conductivity and Hall resistivity of shocked argon: calculations}\label{sec4}

\subsection{Description of our plasma equation-of-state model}\label{subsec35}

The solution of the Rankine-Hugoniot relations requires the knowledge of the equation of state. Estimations of the downstream temperature $T_H$ request $U(\rho,T)$ and $P(\rho,T)$. To this end, we build an equation-of-state model for argon according to the decomposition:
\begin{equation}
\left\{\begin{array}{ccc}
       U(\rho,T)&=&U_c(\rho)+U_{i,\mathrm{th}}(\rho,T)+U_{e,\mathrm{th}}(\rho,T)\\
       P(\rho,T)&=&P_c(\rho)+P_{i,\mathrm{th}}(\rho,T)+P_{e,\mathrm{th}}(\rho,T).
       \end{array}
\right.
\end{equation}
$U_c(\rho)$ and $P_c(\rho)$ denote the 0 K-isotherms, also named ``cold curves''. $U_{e,\mathrm{th}}(\rho,T)$ and $P_{e,\mathrm{th}}(\rho,T)$ are the electronic thermal contributions, obtained by removing the $T=0$ K electronic energies and pressures from the total electronic ones:
\begin{equation}
    \left\{\begin{array}{ccc}
    U_{e,\mathrm{th}}(\rho,T)&=&U_e(\rho,T)-U_e(\rho,0),\\
    P_{e,\mathrm{th}}(\rho,T)&=&P_e(\rho,T)-P_e(\rho,0).
    \end{array}
\right. 
\end{equation}
$U_e(\rho,T)$ and $P_e(\rho,T)$ are calculated with our average-atom code {\sc Paradisio}. The cold contributions $U_c(\rho)$ and $P_c(\rho)$ are the ones of the {\sc Sesame} equation of state {\sc Sesame 5172} of argon \cite{Carpenter2012,Root2022}. {\sc Sesame 5172} incorporates the physics of six theoretical models. It  provides very good agreement with experimental shock data in the very low density range (initial  density: $\rho_0=1.34\times 10^{-3}$ g/cm$^3$) \cite{Carpenter2012,Garcia2017}, thereby justifying our choice of these cold contributions. A more recent {\sc Sesame 5173} model \cite{Root2022} was developed to improve agreement with high pressure Hugoniot (above 90 GPa) as well as with low-temperature data for fluid and solid argon, including phase boundaries, \emph{i.e.}, in areas outside the scope of our study. Finally, the {\sc Ocp} (one-component plasma) model \cite{Hansen1973} is used for the thermal ionic contributions:
\begin{equation}
    \left\{\begin{array}{ccc}
    U_{i,\mathrm{th}}(\rho,T)&=&\rho k_BT+\dfrac{\rho}{3}\Delta U_i(\rho,T)\\
    P_{i,\mathrm{th}}(\rho,T)&=&\dfrac{3}{2}k_B T+\Delta U_i(\rho,T),
    \end{array}
    \right.
\end{equation}
where
\begin{equation}\label{deltae}
    \displaystyle\dfrac{\Delta U_i(\rho,T)}{k_B T}=\min\left(\left[\Gamma^{3/2}\sum_{k=1}^4\dfrac{a_k}{(b_k+\Gamma)^{k/2}}-a_1\Gamma\right],\dfrac{3}{2}\right),
\end{equation}
$\Gamma$ being the usual ionic coupling parameter
\begin{equation}
    \Gamma=\dfrac{{Z^*}^2}{(k_B T)R_{\mathrm{ws}}}.
\end{equation}
The values of the parameters $a_k$ and $b_k$ are given in Table \ref{tab:my_label1}:
\begin{table}[ht!]
    \centering
    \begin{tabular}{c c c}
    \hline 
    $k$ & $a_k$ & $b_k$ \\
    \hline 
    1 & -0.895929 & 4.666486 \\
    2 & 0.11340656 & 13.675411 \\
    3 & -0.90972827 & 1.8905603 \\
    4 & -0.11614773 & 1.0277554 \\
    \hline
    \end{tabular}
    \caption{Parameters entering the {\sc Ocp} plasma model for the ionic contribution to the equation of state (see Eq. (\ref{deltae})).}
    \label{tab:my_label1}
\end{table}

\subsection{Thermodynamic conditions reached in the shock experiments of Shilkin \emph{et al.}}\label{subsec36}

Table \ref{tab:my_label2} gives the thermodynamic conditions $\rho_H$, $T_H$ and $P_H$ obtained from solving the Rankine-Hugoniot relations, and using our EOS model for argon, for each experimental shock velocities $D$ provided by Shilkin \emph{et al.}. Before the initial shocks, argon is assumed to be at ambient temperature $T_0=300$ K and at pressure $P_0=0.4$ MPa \cite{Shilkin2003}. According to our EOS, these conditions imply that the initial argon gas density is $\rho_0=6\times 10^{-3}$ g/cm$^3$.\\

\begin{table}[ht!]
    \centering
    \begin{tabular}{c c c c c}
    \hline 
    $D-V_0$ & $\rho_H$ & $T_H$ & $P_H$ & $V_H-V_0$\\
    (km/s)  & (g/cm$^3$) & (K) & (GPa) & (km/s) \\
    \hline 
    \multicolumn{5}{c}{Principal Hugoniot ($V_0=0$)} \\
    \hline\hline
    2.53 & 2.457 $\times 10^{-2}$ & 5199 & 2.945 $\times 10^{-2}$ & 1.912 \\
    2.56 & 2.461 $\times 10^{-2}$ & 5315 & 3.016 $\times 10^{-2}$ & 1.936 \\
    2.63 & 2.468 $\times 10^{-2}$ & 5541 & 3.153 $\times 10^{-2}$ & 1.980 \\
    2.85 & 2.500 $\times 10^{-2}$ & 6498 & 3.748 $\times 10^{-2}$ & 2.166 \\
    2.91 & 2.510 $\times 10^{-2}$ & 6757 & 3.915 $\times 10^{-2}$ & 2.216 \\
    2.97 & 2.523 $\times 10^{-2}$ & 7000 & 4.077 $\times 10^{-2}$ & 2.264 \\
    3.20 & 2.588 $\times 10^{-2}$ & 7944 & 4.763 $\times 10^{-2}$ & 2.458 \\
    3.27 & 2.613 $\times 10^{-2}$ & 8223 & 4.985 $\times 10^{-2}$ & 2.519 \\
    3.40 & 2.664 $\times 10^{-2}$ & 8737 & 5.418 $\times 10^{-2}$ & 2.634 \\
    3.51 & 2.712 $\times 10^{-2}$ & 9159 & 5.802 $\times 10^{-2}$ & 2.734 \\ 
    \hline 
    \multicolumn{5}{c}{Behind the reflected shock wave, with $B=5$ T} \\
    \multicolumn{5}{c}{($V_0$ : previous $V_H$ values)} \\
    \hline\hline     
    2.53 & 6.357 $\times 10^{-2}$ & 10423 & 1.600 $\times 10^{-1}$ & 0 \\
    2.56 & 6.413 $\times 10^{-2}$ & 10564 & 1.638 $\times 10^{-1}$ & 0 \\
    2.63 & 6.537 $\times 10^{-2}$ & 10897 & 1.728 $\times 10^{-1}$ & 0 \\
    2.85 & 6.945 $\times 10^{-2}$ & 11875 & 2.027 $\times 10^{-1}$ & 0 \\
    2.91 & 7.063 $\times 10^{-2}$ & 12141 & 2.115 $\times 10^{-1}$ & 0 \\
    2.97 & 7.191 $\times 10^{-2}$ & 12407 & 2.210 $\times 10^{-1}$ & 0 \\
    3.20 & 7.753 $\times 10^{-2}$ & 13429 & 2.624 $\times 10^{-1}$ & 0 \\
    3.27 & 7.942 $\times 10^{-2}$ & 13740 & 2.762 $\times 10^{-1}$ & 0 \\\hline
    \end{tabular}
    \caption{First part of the table: principal Hugoniot. The initial matter velocity is $V_0=0$. Second part : after the reflected shock wave, assuming ``frozen-in'' magnetic field lines. The initial matter velocities $V_0$ are the velocities $V_H$ reached on the principal Hugoniot. Since the shock is totally reflected on the obstacle, the downstream mass velocity is the opposite of $V_0$ and therefore $V_H-V_0=0$.}
    \label{tab:my_label2}
\end{table}

\begin{figure}[ht!]
\centering
\includegraphics[scale=0.5]{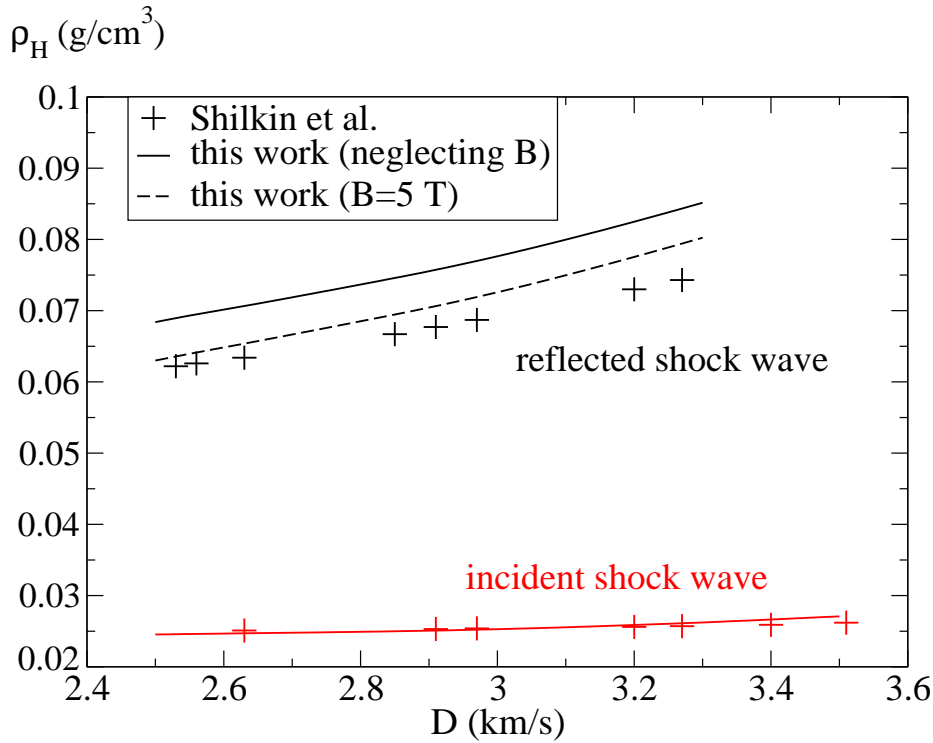}
\caption{\label{fig1} Density versus shock velocity on the principal (incident shock wave) and secondary (reflected shock wave) Hugoniot curves. Comparison between our results without magnetic field, with a magnetic field $B=5 $T, and the experiments of Shilkin \emph{et al.} \cite{Shilkin2003}.}
\end{figure}

\begin{figure}[ht!]
\centering
\includegraphics[scale=0.5]{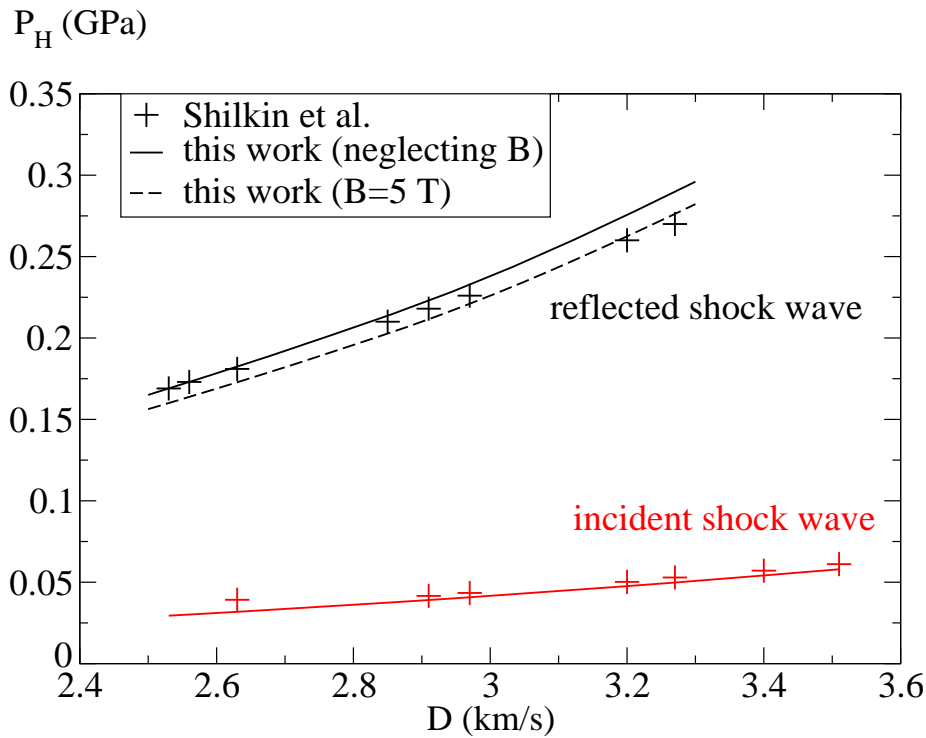}
\caption{\label{fig2} Pressure versus shock velocity on the principal (incident shock wave) and secondary (reflected shock wave) Hugoniot curves. Comparison between our results without magnetic field, with a magnetic field $B=5$ T, and the experimental values \cite{Shilkin2003}.}
\end{figure}

\begin{figure}[ht!]
\centering
\includegraphics[scale=0.5]{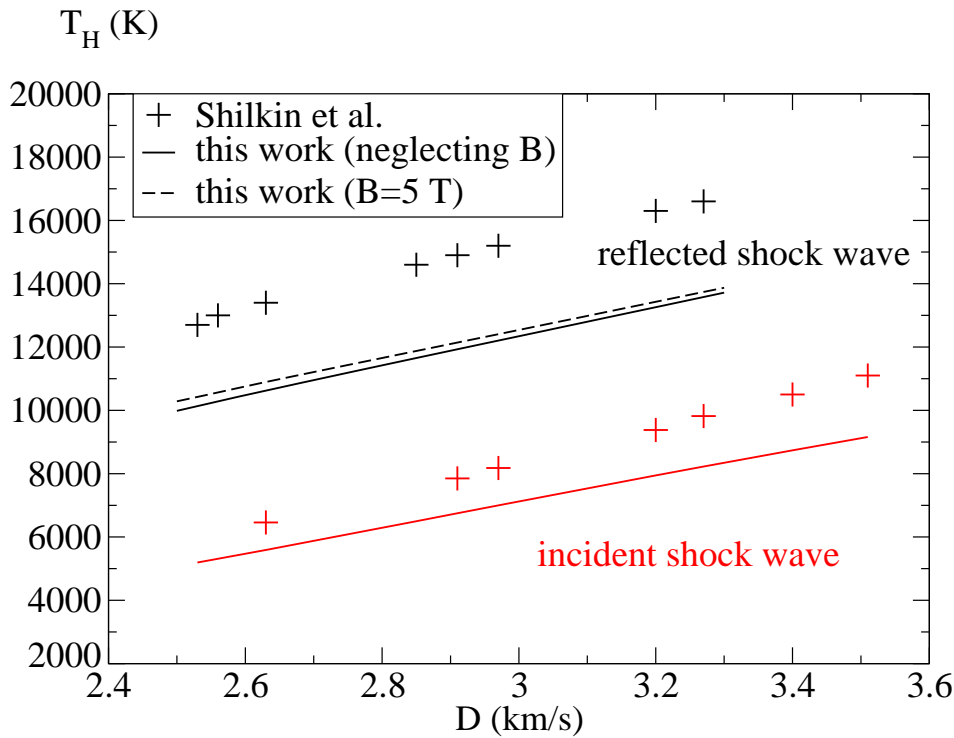}
\caption{\label{fig3} Temperature versus shock velocity on the principal (incident shock wave) and secondary (reflected shock wave) Hugoniot curves. Comparison between our results without magnetic field, with a magnetic field $B=5$ T, and the calculated values of Shilkin \emph{et al.} \cite{Shilkin2003}.}
\end{figure}

Figures \ref{fig1}, \ref{fig2} and \ref{fig3} display respectively the density, temperature and pressure as functions of the shock velocity in the conditions of the shock experiments of Shilkin \emph{et al.} \cite{Shilkin2003}. The crosses represent experimental values, in red for the initial shocks, and in black for the reflected ones. The lines, with the same color code, correspond to our calculations of the conditions reached in these experiments. For the reflected shocks, we present two results, one obtained when the magnetic field is absent in the Rankine-Hugoniot equations (black full line) and the other when taking into account the field $B=5$ T (black dashes). The former case supposes that the reflected shocks do not ionize the argon plasma enough to ensure ``frozen-in'' of magnetic lines, whereas the latter considers that ionization is strong enough to fully get this effect. Our results assuming ``frozen-in'' of the applied $B=5$ T magnetic field present the closest agreement with the experimental densities and pressures. The agreement is particularly improved as concern the densities (see Fig.~\ref{fig1}). The temperature reached downstream the shock waves was not measured by Shilkin \emph{et al.}. The values (crosses) presented in Fig.~\ref{fig3} are theoretical, and noticeably higher than our own theoretical values. These discrepancies are due to different equation-of-state models, the temperature being particularly sensitive to them.

\subsection{Electrical conductivity and Hall coefficient of shocked argon: comparison between experiment and theory}\label{subsec43}

\begin{figure}[ht!]
\centering
\includegraphics[scale=0.5]{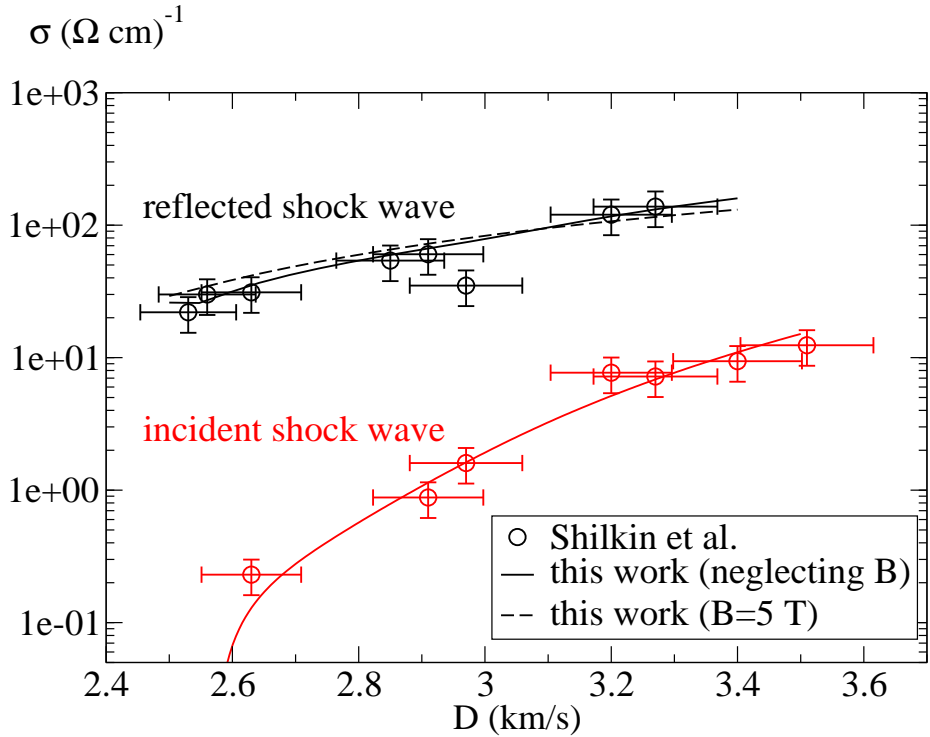}
\caption{\label{fig4} Electrical conductivity versus shock velocity on the principal (incident shock wave) and secondary (reflected shock wave) Hugoniot curves. Comparison between our results without magnetic field, with a magnetic field B = 5 T, and the experiments of Shilkin \emph{et al.} \cite{Shilkin2003}. \label{sigma_Ar}}
\end{figure}

Figure \ref{sigma_Ar} represents conductivity calculations of argon for incident and reflected shock waves, compared to the measurements by Shilkin \emph{et al.}. The calculated conductivities are the inverse of the resistivities obtained with Eq.~(\ref{eta}), using the phase-shifts and mean ionic charges given by the average-atom code {\sc Paradisio} for argon at the densities $\rho_H$ and temperatures $T_H$ given in Table \ref{tab:my_label2}. The color code is the same as in Figs.~\ref{fig1}, \ref{fig2} and \ref{fig3}. We observe global good agreement with experimental values, both for the incident and reflected shock waves. For the latter ones, we present results obtained when the magnetic field is taken into account or not in the shock equations. They differ only slightly, despite fairly different upstream densities (see Fig.~\ref{fig1}).\\

Hall effect in plasmas has been investigated using transport-equation theories, mainly in the non-degenerate limit. These approaches consider interaction between individual species composing the plasma, while our average-atom one describes electrons interacting with others and with mean ions through a mean-field. The most accurate models include electron-electron and electron-neutral atom collisions in addition to electron-ion ones. Table \ref{table3} gives the Hall constant values obtained in the non-degenerate limit according to the collision terms taken into account by the transport-equation models, and compares them with our mean-atom result. More details on these models are given in the following text.\\ 

The electrical and thermal conduction of fully ionized plasma, (\emph{i.e.}, a plasma formed of electrons and ions, with no neutral atoms) has been studied by Spitzer and H\"arm in the classical low density limit within kinetic theory \cite{Spitzer1953}. The linearized Fokker-Planck kinetic equation is solved with a Landau collision integral including both electron-ion (e-i) and electron-electron (e-e) collisions.\\  
Braginskii introduced the magnetic field in this approach, extending it to the Hall effect \cite{Braginskii1965}. An expression has been established for the Hall resistivity  $\eta_\mathrm{Hall}=R_\mathrm{Hall} B$ in terms of powers of $(\omega_c\tau)$
\begin{equation}
    \eta_\mathrm{Hall}=\dfrac{B}{en_e}+\dfrac{m_e}{e^2n_e\tau}\left[\dfrac{(\omega_c\tau)(\alpha_0^{\prime\prime}+\alpha_1^{\prime\prime}(\omega_c\tau)^2)}{\delta_0+\delta_1 (\omega_c\tau)^2+(\omega_c\tau)^ 4}\right].
\end{equation}
At the limit $\omega_c\tau\ll 1$, the Braginskii Hall constant $r_\mathrm{Hall}$ reads
\begin{equation}
    r_\mathrm{Hall}=en_e\times R_\mathrm{Hall}=1+\dfrac{\alpha_0^{\prime\prime}}{\delta_0}.
\end{equation}
The Lorentz plasma is a plasma with highly ionized ions, no neutral atoms, and in which electron-electron collisions can be neglected. For that plasma, Braginskii calculates  $\alpha_0^{\prime\prime}=0.094$ and $\delta_0=0.0961$, yielding $r_\mathrm{Hall}^{Z\gg 1}=1.978$. 
When the atoms are only once ionized: $\alpha_0^{\prime\prime}=0.7796$, $\delta_0=3.7703$, and $r_\mathrm{Hall}^\mathrm{Z=1}=1.207$.\\
Braginskii as well as Spitzer-H\"arm theories are rigorously valid only for fully ionized, (\emph{i.e.}, all atoms are at least ionized one) non-degenerate plasmas.\\
Lee and More's model of transport properties \cite{Lee1984} take into account the electron degeneracy by using the Fermi-Dirac distribution function for the electrons. Boltzmann's equation is solved within the relaxation-time approximation (RTA). Electrical and thermal conductivity, thermoelectric power, and also Hall, Nernst, Ettinghausen and Leduc-Righi coefficients, essential to the study of plasmas in presence of electromagnetic fields, are considered. The transport properties are expressed in computationally simple forms and apply to any electron degeneracy. In the completely non-degenerate limit $(\mu\beta\rightarrow -\infty)$, the Hall constant value is $r_\mathrm{Hall}^{LM}=1.9328$, and is close to Braginskii's one for the Lorentz plasma, which assumes that all atoms are strongly ionized. The ``standard''  $r_\mathrm{Hall}=1$ value for solids is recovered in the totally degenerate limit $\mu\beta\rightarrow \infty$.\\
Stygar, Gerdin and Fehl \cite{Stygar2002} developed a quantum-mechanical approach for the electrical conductivity tensor, for a Lorentz plasma in a weak magnetic field, within the linearized Boltzmann transport approach.   Stygar \emph{et al.} evaluate the Coulomb logarithms in the second Born approximation. They read
\begin{equation}
    \ln \Lambda(v_e)=\left(\ln \chi -\dfrac{1}{2}\right)+\left[\left(\dfrac{2Z^*e^2}{\lambda m_e v_e^2}\right)\left(\ln\chi-\ln 2^{4/3}\right)\right]
\end{equation}
with $\chi=2m_e v_e\lambda/\hbar$, $\lambda=\max(\lambda_D,R_\mathrm{ws})$, 
$\lambda_D$ being the Debye length, given by:
\begin{equation}
    \lambda_D=\left[\left(\dfrac{4\pi n_e e^2}{k_B T}\right)+\left(\dfrac{4\pi Z^* n_e e^2}{k_B T}\right)\right]^{-1/2}.
\end{equation}
$v_e$ denotes the electron velocity. Finally, Stygar \emph{et al.} obtained  the following expression for the Hall constant:
\begin{equation}
    r_\mathrm{Hall}^{SGF}=\dfrac{315\pi}{512}\left(\dfrac{\ln \Lambda(v_{e1})}{\ln\Lambda(v_{e2})}\right)^2,
\end{equation}
with
\begin{equation}
    \begin{gathered}
    v_{e1}=\left(\dfrac{7 k_B T}{m_e}\right)^{1/2},\\
    v_{e2}=\left(\dfrac{10 k_B T}{m_e} \right)^{1/2}.
    \end{gathered}
\end{equation}
$\dfrac{315\pi}{512}\approx 1.9328$, \emph{i.e.}, Lee and More's value for $r_\mathrm{Hall}$ in the non-degenerate limit. Predicted values of the Hall constant applying Lee and More's model and Stygar \emph{et al.}'s one are respectively represented in Fig.~\ref{fig4c} by the black dashes and the black line. \\

The plasma ionization downstream the shock waves generated in argon in the experiments of Shilkin \emph{et al.} is far too weak for the application  of Lee and More as well as Stygar \emph{et al.}'s models, which both assume Lorentz plasmas. These models, and more generally any RTA approach, do not recover Spitzer and H\"arm's result for electrical conductivity in the non-degenerate limit. This is attributed to the fact that e-e collisions, not taken into account in the RTA, grow in importance when the atoms are less ionized and that they can then no more be neglected. Interpolation procedures have been proposed to correct the RTA electrical conductivities \cite{Stygar2002,Fortov2003}, but there is no equivalent for correcting RTA Hall constants.\\
 
Adams \emph{et al.} used an approach based on linear response theory (LRT) within the Zubarev formalism \cite{Zubarev1996,Ropke2013} that allows for a systematic treatment of e-e collisions at any degeneracy \cite{Adams2010}. LRT is a quantum statistical approach based on the grand canonical ensemble, linearized with respect to non-equilibrium perturbations such as external fields \cite{Ropke1988,Ropke2013,Reinholz2015}. LRT takes into account all interactions, including e-e ones, through equilibrium force-force correlation functions. Electron-neutral-atom (e-n) collisions are also taken into account. Transport coefficients are calculated using a converging expansion in terms of so called generalized moments. When e-e collisions are neglected in the theory, Adams \emph{et al.} recover the $r_\mathrm{Hall}=1.9328$ RTA value in the non-degenerate limit, and obtain $r_\mathrm{Hall}=1.1994$ when e-e collisions are accounted for. The latter value is very close to the one $r_\mathrm{Hall}^{Z=1}=1.207$ calculated by Braginskii for atoms ionized once. When e-n collisions are taken into account, the Hall constant is enhanced up to $r_\mathrm{Hall}\approx 1.5$ for the less degenerate argon plasmas, according to Ref. \cite{Adams2007b}. Since this work, the authors presented in \cite{Adams2010} an extension of LRT to include the effects of an external magnetic field, which results in a value only slightly higher than the standard $r_\mathrm{Hall}=1$ Hall constant.\\

\begin{figure}[ht!]
\centering
\includegraphics[scale=0.5]{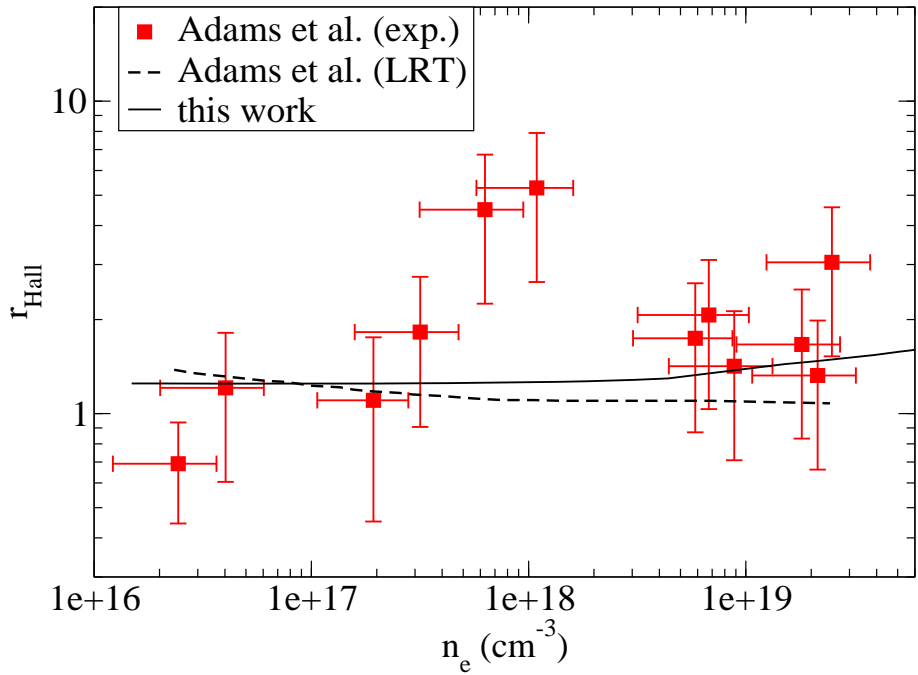}
\caption{\label{fig4b} $r_\mathrm{Hall}=R_\mathrm{Hall}\times en_e$ as a function of electron density $n_e$. Red squares and black dashes: the experimental and theoretical (using LRT) values of Adams \emph{et al.} \cite{Adams2007b}. Black full line: our results.}
\end{figure}

\begin{figure}[ht!]
\centering
\includegraphics[scale=0.5]{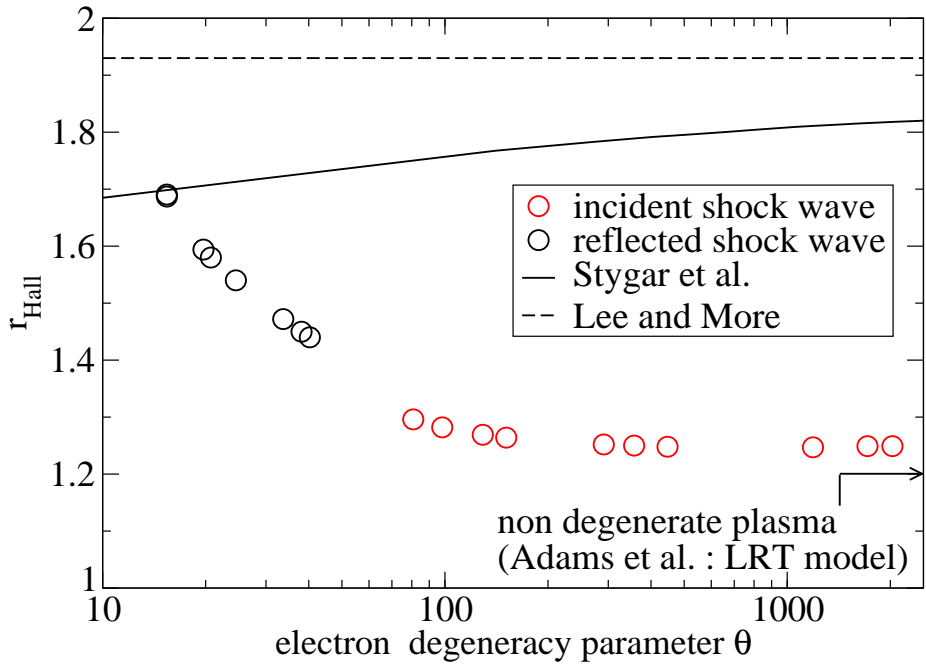}
\caption{\label{fig4c}$r_\mathrm{Hall}$ versus electron degeneracy parameter $\Theta$. Red circles : our results, in conditions reached downstream the principal shock waves. Black circles: our results, behind the reflected shocks. Black dashes: Lee and More model \cite{Lee1984}. Black line: Stygar \emph{et al.} model. The arrow points to the value predicted by Adams \emph{et al.} in the totally non-degenerate limit when electron-electron collisions are taken into account in collision integrals, for the fully ionized $Z=1$ plasma \cite{Adams2010}. }
\end{figure}

Fig.~\ref{fig4b} compares our calculated $r_\mathrm{Hall}$ constants (black line) with the experimental ones (red squares) deduced by measured Hall voltages by Shilkin \emph{et al.}. Hall voltage is proportional to $R_\mathrm{Hall}$, and the experimental $r_\mathrm{Hall}$ are obtained using theoretical electron densities $n_e$ from SAHA IV code. The figure also presents the theoretical values obtained by Adams \emph{et al.} using the linear response theory approach \cite{Reinholz2015}.\\

Our Ziman average-atom approach takes into account interactions between mean ions and electrons, and the interactions between electrons through their total charge density and through the exchange-correlation potential. At high degeneracy parameters $\Theta$, we calculate values for the Hall constant $r_\mathrm{Hall}$ close to Adams \emph{et al.}'s LRT and Braginskii's one for a low density plasma composed of ions with the lowest possible charge $Z=1$, no neutral atoms and electrons, where the e-e collisions between are considered (see Fig.~\ref{fig4c}, where the arrow points on that value).\\ 
As electron degeneracy increases, (\emph{i.e.}, as $\Theta$ decreases), Adams \emph{et al.} predict, within the LRT approach, the decrease of $r_\mathrm{Hall}$ in shocked argon in Shilkin \emph{et al.} experiments as shown by the black dashes in Fig.~\ref{fig4b}  (the degeneracy parameter $\Theta$ varies inversely to $n_e$). At the opposite, we obtain increased  $r_\mathrm{Hall}$ values as electron degeneracy rises, getting closer to the RTA values, which are considered as becoming more relevant as degeneracy is stronger. Oppositely, at the highest electron density obtained in the experiments, for which we estimate degeneracy parameters of the order of 15, Adams \emph{et al.} calculate  $r_\mathrm{Hall}$ values, decreasing with $\Theta$, and becoming very close to the ``standard'' $r_\mathrm{Hall}=1$ value expected for highly degenerate plasmas ($\Theta\ll 1$) and for solids.\\ 

As in the non-degenerate case, we summarize, in table \ref{table4}, the results obtained with transport-equation methods according to the included collision terms, in the case of the partially degenerate argon plasma ($\Theta=20$), and compare them to our average-atom result.\\

In next section, we are going to take a closer look at our comparisons with the LRT calculations of Adams \emph{et al.}.

\begin{table}[!ht]
\centering
    \begin{tabular}{c | c c c | c }
    \hline
    \multicolumn{5}{c}{{\sc non-degenerate limit}}\\
    \hline
    model & \multicolumn{3}{c|}{included collision terms}  & Hall constant \\
          & e-i & \hspace{5mm} e-e & \hspace{5mm}e-n & $r_\mathrm{Hall}$ \\
    \hline 
    Braginskii \cite{Braginskii1965} & {\Huge $\times$} & \hspace{5mm}{\Huge $\times$} & & 1.207\\
    Lee and More \cite{Lee1984} & {\Huge $\times$} & & & 1.9328\\
    Stygar \emph{et al.} \cite{Stygar2002}  & {\Huge $\times$} & & & 1.9328\\
    LRT 1 \cite{Adams2007b,Adams2010} & {\Huge $\times$} &  &  & 1.933\\    
    LRT 2 \cite{Adams2007b,Adams2010} & {\Huge $\times$} & \hspace{5mm}{\Huge $\times$} &  & 1.199\\
    LRT 3 \cite{Adams2007b} & {\Huge $\times$} & \hspace{5mm}{\Huge $\times$} & \hspace{5mm}{\Huge $\times$} & $\simeq$ 1.5\\
    \hline
    our work & \multicolumn{3}{c|}{mean collision time} & 1.25\\
    $(\Theta=1000)$ & \multicolumn{3}{c|}{(AA and T-matrix)} & \\
    \hline
    \end{tabular}
    \caption{We compare our calculated Hall constant value to the ones obtained with transport-equation approaches in the case of non-degenerate electrons. The crosses indicate the collision terms taken into account in different models. Our average-atom (AA) result is close to the Braginskii and the LRT (linear response theory) numbered 2 ones accounting for direct electron-electron interactions, as well as to the LRT 3 one that also includes electron-neutral atom collisions.}
    \label{table3}
\end{table}

\begin{table}[!ht]
\centering
    \begin{tabular}{c | c c c | c }
    \hline
    \multicolumn{5}{c}{{\sc partial electron degeneracy} $\Theta=20$}\\
    \hline
    model & \multicolumn{3}{c|}{included collision terms}  & Hall constant \\
          & e-i & \hspace{5mm} e-e & \hspace{5mm}e-n & $r_\mathrm{Hall}$ \\
    \hline 
    Lee and More \cite{Lee1984} & {\Huge $\times$} & & & 1.93\\
    Stygar \emph{et al.} \cite{Stygar2002}  & {\Huge $\times$} & & & 1.69\\    
    LRT  \cite{Adams2007b,Adams2010} & {\Huge $\times$} & \hspace{5mm}{\Huge $\times$} & \hspace{5mm}{\Huge $\times$} & $\simeq$ 1.\\
    \hline
    our work & \multicolumn{3}{c|}{mean collision time} & 1.69\\
     & \multicolumn{3}{c|}{(AA and T-matrix)} & \\
    \hline
    \end{tabular}
    \caption{We compare our calculated Hall constant value to the ones obtained with transport equation approaches in the case of partial electron degeneracy parameter $\Theta=20$. Our result is close to the ones of Lee and More and of Stygar \emph{et al.}, which are becoming more relevant since electron exchange-correlation effects are becoming increasingly important compared with e-e direct collisions. In the text, we suggest a possible explanation for the fact that the LRT is already tending towards the expected $r_\mathrm{Hall}=1$ value for the fully degenerate case.}
    \label{table4}
\end{table}

\section{Discussion: average-atom versus LRT approaches for the Hall constant}\label{sec5}

For solids, as well as for low density hot plasmas, the Hall constant has the so-called ``standard'' value $r_\mathrm{Hall}=1$. Our calculated values in the conditions of the experiments of Shilkin \emph{et al.} are significantly different from the ones obtained by Adams \emph{et al.} \cite{Adams2007,Adams2007b,Adams2007c,Adams2010} using LRT approach within Zubarev's method, which raises questions.\\ 

The LRT approach takes into account electron-ion, electron-neutral and electron-electron collisions, enabling for a complete description of partially ionized plasmas, for which the value of $r_\mathrm{Hall}$ is unknown. One limitation to its use could be the difficulty to build cross-sections for the scattering of electrons by neutral atoms.  Because of the lack of theoretical electron-atom cross-sections for argon at the date of their work, Adams \emph{et al.} \cite{Adams2007b,Adams2010}  used experimental data, obtained for argon at ambient temperature. The method used for the calculation of the plasma composition (density of neutrals, ions and electrons) also introduces some uncertainty. In the case of the largest argon densities in the experiments of Shilkin \emph{et al.}, Adams \emph{et al.} reported as much as 40\% differences in the theoretical electron densities obtained according to whether the {\sc Saha IV} code of Gryaznov \cite{Gryaznov1973} or the {\sc Comptra} program \cite{Kuhlbrodt2005} is used for that purpose \cite{Adams2007b} (both codes are based on similar chemical pictures for the plasma equations of state, but use different thermodynamical models).\\

The average-atom approach used in the present work presents its own difficulties. First, the separation between bound and free electrons may be problematic \cite{Wetta2022,Wetta2023}. However, this does not concern argon in the density and temperature ranges reached in the experiments of Shilkin \emph{et al.}. The bound-free separation is unambiguous, and all possible definitions for the mean ion charge $Z^*$ \cite{Wetta2023} converge to the same value. It remains the question of properly  accounting for e-n and e-e interactions with average-atom methods.\\

\begin{figure}[ht!]
\centering
\includegraphics[scale=0.5]{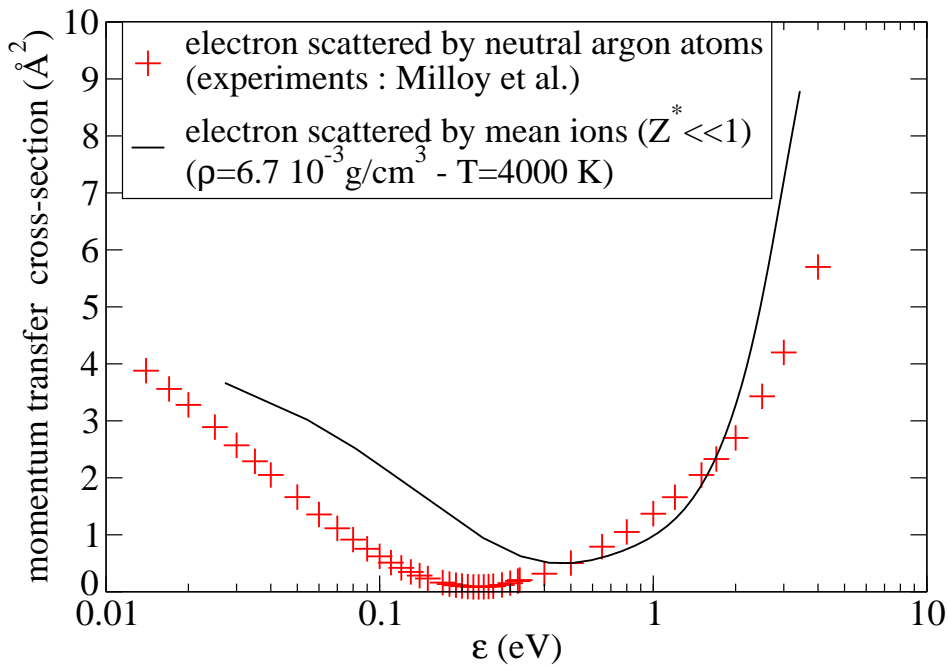}
\caption{\label{fig5}Momentum transfer cross-section for electron- neutral argon collisions. Red crosses: experiments of Milloy \emph{et al.} \cite{Milloy1977}. Black line: the average-atom code is used to calculate the cross-section for an electron scattered by an almost neutral argon atom (case of argon at density $\rho=6.7\times 10^{-3}$ g/cm$^3$ and temperature $T=4000$ K, for which $Z^*<10^{-5}$). }
\end{figure}

Let us start looking at the way that average-atom methods handle the scattering of electrons by neutral atoms. In the experiments of Shilkin \emph{et al.}, kinetic models consider that argon plasmas are composed, outside the electrons, by neutral and ionized argon atoms, and use two distinct approaches, on the one hand for the e-n collision times and on the other for all e-i ones. The average-atom approach used in the present work avoids the problem of distinguishing ions and neutral atoms, which are replaced by identical ions with the same mean charge $Z^*$. In the considered experiments, the neutral atoms are about $10^2$ up to $10^6$ times more numerous than ionized argon atoms. Thus, the mean ion is almost a neutral argon atom, and the average-atom model actually provides an e-n collision time that extrapolates the average-atom e-i collisions time to the limit $Z^*\rightarrow 0$. Posterior to Adams \emph{et al.}' works, Quan \emph{et al.} derived e-n and e-i model potentials with the aim to build theoretical e-n and e-i scattering cross-sections \cite{Quan2020}. For e-n scattering the model potential reads
\begin{equation}
V(r)=V_s(r)+V_p(r)+V_x(r),   
\end{equation}
where $V_s(r)$ is the sum of the electron-nucleus Coulomb potential and of the free electron-bound electrons Coulomb interactions, $V_x(r)$ an exchange potential, and $V_p(r)$ a polarisation potential. The e-i model potential only differs from the e-n one by a screening factor $e^{-r/r_D}$, $r_D=\sqrt{k_B T/4\pi n_e}$ being the Debye screening length, which applies to the Coulomb term $V_s(r)$. Quan \emph{et al.}'s e-i model potential tends then to e-n one in the limit $Z^*\rightarrow 0$ (since $n_e\rightarrow 0$), and the e-n scattering cross-section appears as the $Z^*\rightarrow 0$ limit of the e-i one, as in the average-atom approach.\\ 
Average-atom effective potential includes the same two Coulomb contributions and electron exchange potential too. Polarisability does not appear explicitly, but is, in some way, present through electron exchange and correlations. The problem is whether this is sufficient to reproduce the experimental cross-sections, given that the polarization potential makes a significant contribution to the model of Quan \emph{et al.} \cite{Quan2020}. Unfortunately, we were not able to calculate the cross-section for the scattering of electrons by almost neutral atoms for argon at ambient temperature with our average-atom code. We only obtained converged results for somewhat higher densities and temperatures. Figure \ref{fig5} compares the average-atom momentum transfer cross-section for electron collisions with almost neutral atoms obtained for $\rho=6.7\times 10^{-3}$ g/cm$^3$ and $T=4000$ K (represented by the black line), to the experimental cross-section for electron-neutral atom collisions measured for argon at ambient temperature by Milloy \emph{et al.} (red crosses). Qualitatively, the average-atom calculation is in fairly good agreement with the experiments. In particular, a Ramsauer-Townsend like minimum is also predicted with the average-atom approach, albeit at a somewhat higher collision energy than experimentally. One must also keep in mind that the actual plasma temperatures are about $10^4$ K in the experiments. For this reason, we think that our average-atom approach, which allows for electron density and temperature changes, is relevant for e-n interactions, despite the fact that the neutral atoms are approximated by ions carrying low charges $Z^*$.\\

To explain any discrepancies with other theoretical approaches, like the  LRT one, it remains the possibility that the average-atom models do not properly take into account the e-e interactions. The inclusion of e-e direct collisions in Density-Functional-Theory (DFT) approaches for transport properties is subject of discussions in the literature. In a paper published in 2006 \cite{Dharma-wardana2006}, Dharma-wardana advances strong arguments in favor of the DFT-based approaches for e-e interactions. Indeed, he points out that the electron current is conserved under e-e interactions, since the electron current operator commutes fully with the e-e interaction Hamiltonian, and that this holds at any electron degeneracy. The e-e interactions only contribute indirectly to the resistivity through the e-i effective potential. Dharma-wardana develops these arguments further in a very recent article (see section II in the Supplemental Material of Ref. \cite{Dharma-wardana2024}). The counter-argument putted forward against DFT-based approaches is that the latter consider the electrons  as an aggregate, through their total charge density, rather than as individuals interacting with each other like it is done in kinetic theories such as the Boltzmann equation \cite{Reinholz2015,Desjarlais2017}. It is however expected that both DFT based methods and kinetic approaches will converge as electron degeneracy grows, resulting from the increasing compensation of e-e interactions by the exchange-correlation potential.\\
 
The non-degenerate limit $\Theta\rightarrow\infty$ is the most largely studied by kinetic methods, mainly for totally ionized plasmas, in which all ions are at least ionized once. Braginskii's value $r_\mathrm{Hall}=1.207$ when all atoms are ionized once is confirmed by Adams {et al.}'s LRT value $r_\mathrm{Hall}=1.199$ obtained assuming the same plasma composition. Both approaches include e-e direct interactions. When the latter are neglected, $r_\mathrm{Hall}=1.9328$  \cite{Lee1984,Stygar2002}, which is clearly higher. In the electron degeneracy range $100\lesssim \Theta \lesssim 2\,10^3$ corresponding to Shilkin \emph{et al.}'s experiments in incident shocks, in which the mean ion charge $Z^*$ remains small, our average-atom approach yields the value $r_\mathrm{Hall}\approx1.25$  close to the one expected when e-e interactions are properly taken into account. This result backs up the arguments putted forward in the literature in favor of the suitability of DFT-based methods for the study of transport properties.\\

But, accounting within LRT for scattering of electron by neutral atoms besides a low concentration of atoms ionized once, Adams \emph{et al.} obtained very different results, presented in two articles \cite{Adams2007b} and \cite{Adams2007c}. In both works, e-n scattering times are derived from the experimental scattering-cross sections measured at ambient temperature presented in Fig.~\ref{fig5}. In the second paper, the LRT was extended to include explicitly the magnetic field, which was not the case in the first one. Within the extended LRT, the $B=5$ T field applied by Shilkin \emph{et al.} was found to reduce the theoretical $r_\mathrm{Hall}$ value (see the Fig.~4 in \cite{Adams2007c}), whereas, using the non-modified LRT approach, the value is raised up to $r_\mathrm{Hall}\approx 1.5$ in the less degenerate cases. In the modified LRT approach, transport coefficients are written as ratios of polynomials in $X=(\omega_c\tau_0)^2$, where $\tau_0$ is given by the one-moment LRT mean scattering time. The authors observe that, for weakly coupled plasmas, even magnetic fields below  $B=5$ T impact the values of transport coefficients (including electrical, thermal conductivities, as well as Hall constant). \\
As electron degeneracy grows, (\emph{i.e.}, as $\Theta$ diminishes), we observed, within our average-atom approach, a strong dependency of the Hall constant $r_\mathrm{Hall}$ on the electron density and the temperature. This is clearly not predicted by Adams \emph{et al.}, which found that the Hall constant value tends rapidly towards $r_\mathrm{Hall}=1$, as expected for degenerate plasmas, whether or not standard or extended LRT model was used. A possible source for the observed discrepancy between LRT results and ours is the use, within LRT, of experimental ambient temperature scattering cross-section for e-n scattering while the temperatures exceed $10^4$ K. Indeed, it can be easily shown that, using the same cross-section regardless of the increase of temperature, $r_\mathrm{Hall}$ decreases as the $T$ grows.\\
Using the following relation between the cross-section $Q_\mathrm{en}$ for scattering of electron by neutral atoms and the collision time $\tau_\mathrm{en}$ 
\begin{equation}
\hbar k\,n_n\,Q_\mathrm{en}(\epsilon)=\dfrac{1}{\tau_\mathrm{en}(\epsilon)},    
\end{equation}
yields, at temperature $T_0$ 
\begin{equation}
r_\mathrm{Hall}^0 = (3\pi^2 n_e)\times \dfrac{\int_0^\infty \dfrac{k^3}{(k Q_\mathrm{en})^2}\left(-\dfrac{\partial f}{\partial\epsilon} \right)_0 d\epsilon}{\left[\int_0^\infty \dfrac{k^3}{(k Q_\mathrm{en})}\left(-\dfrac{\partial f}{\partial\epsilon} \right)_0 d\epsilon \right]^2}.   
\end{equation}
In the temperature and density conditions considered in this work: $-\left(\dfrac{\partial f}{\partial\epsilon}\right)_0\approx \beta_0 e^{-\beta_0(\epsilon-\mu_0)}$. Let us increase the temperature by a small quantity $\delta T\ll T_0$. At first order of the expansion in $\delta T/T_0$, the resulting variation in the derivative of the Fermi-Dirac distribution function reads
\begin{align}
\left(-\dfrac{\partial f}{\partial\epsilon} \right)&\approx  \left(-\dfrac{\partial f}{\partial\epsilon}\right)_0\,e^{\beta_0 \frac{\delta T}{T_0}(\epsilon-\mu_0)}\nonumber\\
&\approx  \left(-\dfrac{\partial f}{\partial\epsilon}\right)_0\,\left(1+\beta_0 \dfrac{\delta T}{T_0}(\epsilon-\mu_0)\right),  
\end{align}
and the Hall constant at $T=T_0+\delta T$ is
\begin{align}
r_\mathrm{Hall}\approx & r_\mathrm{Hall}^0 \Bigg\{1+\dfrac{\delta T}{T_0}\Big[1+\beta_0  \dfrac{\int_0^\infty \dfrac{k^3(\epsilon-\mu_0)}{(k Q_\mathrm{en})^2}\left(-\dfrac{\partial f}{\partial\epsilon} \right)_0 d\epsilon}{\int_0^\infty \dfrac{k^3}{(k Q_\mathrm{en})^2}\left(-\dfrac{\partial f}{\partial\epsilon} \right)_0 d\epsilon} \nonumber     \\
&-2\beta_0 \dfrac{\int_0^\infty \dfrac{k^3(\epsilon-\mu_0)}{(k Q_\mathrm{en})}\left(-\dfrac{\partial f}{\partial\epsilon} \right)_0 d\epsilon}{\int_0^\infty \dfrac{k^3}{(k Q_\mathrm{en})}\left(-\dfrac{\partial f}{\partial\epsilon} \right)_0 d\epsilon}\Big]\Bigg\}.
\end{align}
In the thermodynamical conditions reached in the experiments of Shilkin \emph{et al.}, the chemical potential has a large negative value $-9\lesssim \mu_0\lesssim -4.5$ (atomic units), and only the lowest energies $\epsilon$ contribute to the integrals. Therefore, taking $(\epsilon-\mu_0)\approx -\mu_0$, the previous equation is simplified to
\begin{equation}
r_\mathrm{Hall}\approx r_\mathrm{Hall}^0\,\left(1+\dfrac{\delta T}{T_0}\left[1-\beta_0\mu_0\right] \right),    
\end{equation}
and finally, using the relation, valid for electron degeneracy parameters $\Theta>1$
\begin{equation}
e^{\beta_0\mu_0}=\dfrac{4}{3\sqrt{\pi}}\Theta_0^{-3/2},    
\end{equation}
one gets, after having replaced $\ln\left[4/(3\sqrt{\pi})\right]$ by its numerical value
\begin{equation}
\dfrac{\delta r_\mathrm{Hall}}{r_\mathrm{Hall}^0}\approx -\dfrac{\delta T}{T_0} \left(0.285+\dfrac{3}{2}\ln \Theta_0 \right).   
\end{equation}
Therefore, \emph{if one uses the same experimental cross-section for the scattering of electrons by neutral atoms regardless of temperature changes}, the calculated Hall constant decreases when temperature grows.\\

\section{Conclusion}\label{sec6}

We presented calculations of the resistivity of shocked argon within Ziman formalism combined with relativistic quantum average-atom method. We have compared our results with measurements performed in experiments involving both incident and reflected shock waves, in presence of a magnetic field of 5 Tesla. Beyond the experimental electric conductivities, the other important objective of these experiments was to measure the Hall resistivity,  in view of deducing the experimental electron density, or, equivalently, the mean ionic charge.\\
The average-atom code {\sc Paradisio} was used for the calculation of the equation of state for argon, as well as for the scattering phase-shifts needed for the scattering amplitude of electrons by the mean ions, and for the mean ion charge $Z^*$. We took into account the magnetic field in the Rankine-Hugoniot relations and derived, starting from the Boltzmann equation, the resistivity tensor in terms of the mean electron-ion collision time $\tau(\epsilon)=\Lambda(\epsilon)/v$ (see section \ref{subsec42}), \emph{i.e.}, the inverse of the collision frequency used in the Ziman resistivity formula.\\

It turns out that the effect of a 5 Tesla magnetic field on the calculation of electrical conductivity is rather limited in the conditions of the experiments. This also justifies the small magnetic field assumption $\omega_c\tau\ll 1$ made for the derivation of the resistivity tensor $\overline{\overline{\eta}}$. The off-diagonal element $\eta_{12}$ giving the Hall resistivity reads then $R_\mathrm{Hall} B=r_\mathrm{Hall}B/(e n_e)$, where the dimensionless Hall constant $r_\mathrm{Hall}$ is the ratio $r_\mathrm{Hall}=\dfrac{\langle\tau^2\rangle}{\langle\tau\rangle ^2}$.\\
We presented Hall constant $r_\mathrm{Hall}$ calculations based on the use of the average-atom code {\sc Paradisio} for the relaxation time $\tau(\epsilon)$, in the conditions reached in the shock experiments of Shilkin \emph{et al.} carried out on argon. In our approach, $\tau(\epsilon)=\Lambda(\epsilon)/v$ is the inverse of the mean electron-ion collision frequency used for the Ziman resistivity calculation. We compared our results to experimental values derived from the Hall voltage  measurements by Shilkin \emph{et al.}, as well as to theoretical ones from Adams \emph{et al.}, based on the quantum statistical linear-relaxation-time approach within the Zubarev formalism.\\
Both sets of results are in the (large) experimental error bars, but within our approach, $r_\mathrm{Hall}$ values rise with electron densities, and are closer to the central experimental values. Our results are in good agreement with Adams \emph{et al.}'s ones in the case of the less degenerate plasmas, for which the relevance of density-functional-theory based models is nevertheless questionable. A growing discrepancy with Adams \emph{et al.} appears as both electron density and temperature rise, which we explain by the fact that Adams \emph{et al.} used ambient temperature experimental scattering cross-section for electron scattering by neutral atoms when the actual temperatures exceed $10^4$ K.

\bibliography{jrnlabbr,refs-noble-gases}

\end{document}